\documentclass[twocolumn,aps,prl,showpacs,amsmath,superscriptaddress,longbibliography,notitlepage]{revtex4-2}
\usepackage{bbm}
\usepackage{}
\usepackage{amssymb}
\usepackage{mathrsfs}
\usepackage{graphicx}
\usepackage{float}
\usepackage[caption=false]{subfig}
\usepackage{verbatim}
\usepackage{txfonts}
\usepackage{epstopdf}
\usepackage[normalem]{ulem}
\usepackage{xcolor}
\usepackage{bm}
\usepackage[utf8]{inputenc}
\usepackage[pdftex,colorlinks,linkcolor={red!75!black},citecolor={red!75!black},urlcolor={red!75!black}]{hyperref}
\usepackage{multirow}
\usepackage{pgfplots}
\usetikzlibrary{patterns,arrows,positioning,calc,fadings,shapes,decorations.markings}

\newcommand{\clr}{\color{red!75!black}}

\def\ket#1{\vert #1 \rangle}
\def\bra#1{\langle #1 \vert}

\begin{document}

\title{Intrinsic-perturbation induced anomalous higher-order boundary states in non-Hermitian systems}

\author{Hui-Qiang Liang}
\affiliation{Department of Physics, Shandong University, Jinan 250100, China}
\author{Zuxuan Ou}
\affiliation{Guangdong Provincial Key Laboratory of Quantum Metrology and Sensing $\&$ School of Physics and Astronomy, Sun Yat-Sen University (Zhuhai Campus), Zhuhai 519082, China}
\author{Linhu Li}\email{lilinhu@quantumsc.cn}
\affiliation{Quantum Science Center of Guangdong-Hong Kong-Macao Greater Bay Area (Guangdong), Shenzhen, China}
\author{Guo-Fu Xu}\email{xgf@sdu.edu.cn}
\affiliation{Department of Physics, Shandong University, Jinan 250100, China}

\date{\today}

\begin{abstract}
The behavior of higher-order boundary states in non-Hermitian systems is elusive and thereby finding the mechanism behind these states is both essential and significant. Here, we uncover a novel mechanism that induces anomalous higher-order boundary states. The mechanism originates from the sensitivity of the non-normal boundary Hamiltonian to intrinsic perturbations, where intrinsic perturbations here refer to the influence of the bulk on the topological boundaries. Based on the mechanism, we reveal a new kind of phase transition, i.e., the transition between hybrid skin-topological states and scale-free topological boundary states. We also find that scale-free topological boundary states exhibit size-dependent spectra, influencing the existence of higher-order topological boundary states. Unlike conventional hybrid skin-topological states or higher-order non-Hermitian skin effect, the above two kinds of anomalous higher-order boundary states exhibit size-dependent characteristics. Our work opens a new horizon for the control of higher-order boundary states and topological properties of non-Hermitian systems.
\end{abstract}
\maketitle

\textit{\it \clr Introduction.}---
Higher-order boundary states generalize the concept of conventional boundary states by manifesting in codimensions greater than one, such as corner or hinge states in two- and three-dimensional systems~\cite{benalcazar_quantized_2017,schindler2018higher,denner_exceptional_2021}. Of late, investigations into higher-order boundary states have extended to systems described by non-Hermitian Hamiltonians~\cite{liu2019second,lee2019hybrid,luo2019higher,li2020topological,okugawa2020second,kawabata_higher-order_2020,denner2021exceptional,zhang_observation_2021,gao_non-hermitian_2021,zou_observation_2021,palacios2021guided,li2022gain,zhang2022universal,li2023enhancement,nakamura_universal_2023,okuma2023non,lin2023topological,zhu_brief_2023,zhu_higher_2023,sun2024photonic,liu2024localization,wu2024spin,liu2024higher-order}.
In fact, the study of higher-order boundary states in non-Hermitian systems has emerged as a frontier in condensed matter physics, bridging the realms of topology and non-Hermitian physics~\cite{okuma2023non,lin2023topological}.
However, as spatial dimensions increase, characterizing the behavior of higher-order boundary states in non-Hermitian systems become dramatically difficult due to the intricate interplay between multiple factors, such as sophisticated topology~\cite{schnyder2008classification,teo_topological_2010,ryu2010topological,chiu2016classification}, geometry~\cite{zhang2022universal}, and non-Hermiticity, where non-Hermiticity includes complex energy spectra, boundary conditions, exceptional points, and the non-Hermitian skin effect (NHSE)~\cite{yao2018edge,yokomizo2019non,liu2019second,lee2019hybrid,luo2019higher,li2020topological,
denner2021exceptional,gou2021exact,li2022gain,zhang2022universal,li2023enhancement,
nakamura_universal_2023,okuma2023non,lin2023topological,zhu_brief_2023}. This makes the behavior of higher-order boundary states in non-Hermitian systems elusive. Thus, to control these states, finding the mechanism that can elucidate the behavior of these states is both essential and significant. But meanwhile, uncovering such a mechanism is full of difficulty because, as mentioned above, the behavior of these states is relevant to many factors.

In this work, we uncover a novel mechanism that modulates the localization behavior of higher-order boundary states, and governs the existence of higher-order topological boundary states. The mechanism is based on the sensitivity of the non-normal boundary Hamiltonian to intrinsic perturbations. Specifically, the affection of the bulk on the topological boundaries, whose information is encoded in the Green’s function of the bulk and which is exceedingly small compared to the other physical quantities, acts as the perturbation to the topological boundaries, and our mechanism origins from the sensitivity of the non-normal boundary Hamiltonian to such an intrinsic perturbation. Based on the mechanism, we reveal a new kind of phase transition. That is, we find that the boundary spectra and the state localization in a non-Hermitian system depend on the geometric aspect ratio of the system, and a certain geometric aspect ratio separates two distinct phases: one phase is hybrid skin-topological states~\cite{lee2019hybrid,zhu_brief_2023,lin2023topological} whose corresponding spectra are line-like; the other phase is scale-free topological boundary states whose corresponding spectra are loop-like or area-like. Based on the mechanism, we also find that the spectra of the scale-free topological boundary states are size-dependent, which affects the existence of the higher-order topological boundary states. As a prototypical example of the size-dependent feature, we introduce a two-dimensional (2D) Benalcazar-Bernevig-Hughes (BBH) model with irreciprocal hopping boundaries that can exhibit size-dependent corner states. Unlike conventional skin-localized topological boundary states ~\cite{lee2019hybrid,liu2019second,denner2021exceptional,okugawa2020second,kawabata_higher-order_2020,zhu_brief_2023}, which exhibit skin localization and are size-independent, the anomalous higher-order boundary states induced by this mechanism display size-dependent characteristics and no longer exhibit skin localization.

\begin{figure}[htb]
\includegraphics[width=0.8\linewidth]{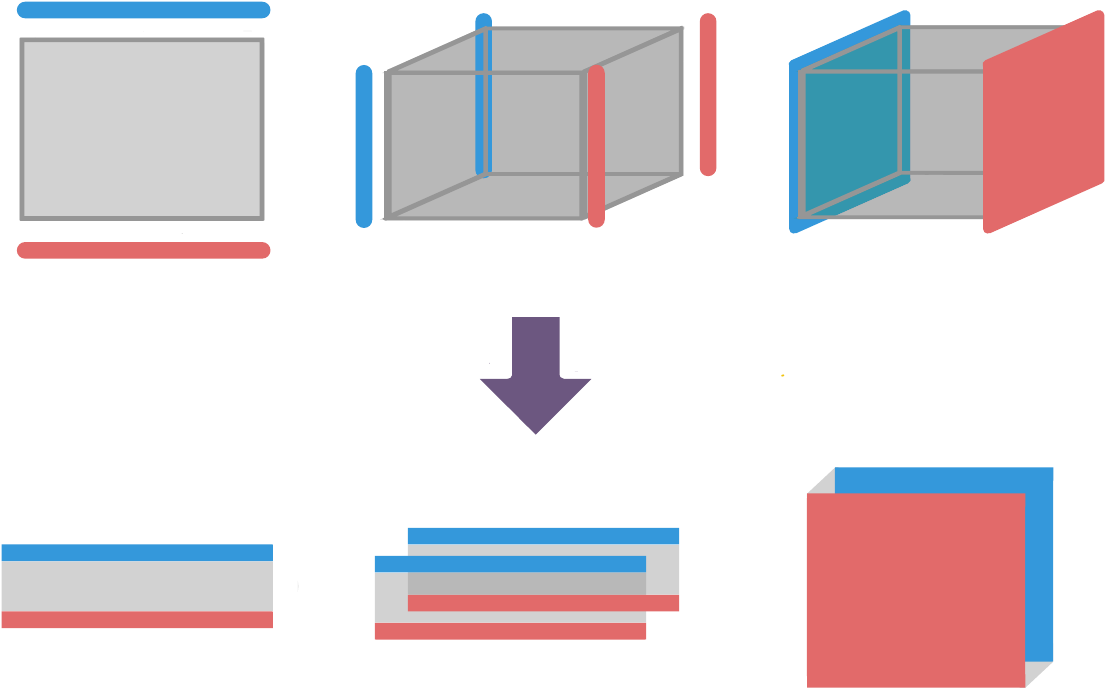}
\caption{The sketch of the mechanism. The red and blue bars/surfaces denote the corresponding physical edges of topological boundary states. While the upper shows real non-Hermitian systems, the bottom shows the corresponding effective boundary model, where the gray parts in the upper represent the bulks of the systems and the gray parts in the bottom represent the affections of the bulks on topological boundaries.}
\label{sketch}
\end{figure}

\textit{\it \clr The mechanism.}---
To convey the insight behind our mechanism, we first lay out a formalism describing the affection of the bulk on the boundaries. Consider a general $d$-dimensional ($d\geq 2$) non-Hermitian lattice, under open boundary conditions (OBCs) along $x$ direction and periodic boundary conditions (PBCs) along the other directions. {The system hosts topological boundary states localized at the top and bottom edges (along $x$) due to a non-trivial bulk topology characterized by the topological invariant~\citep{kawabata_symmetry_2019}.  Without loss of generality, we consider the full Hamiltonian of this system, including the bulk and boundary Hamiltonians, is non-Hermitian.} The Hamiltonian can be written as
\begin{equation}
H = \begin{pmatrix}
H_\text{boundary}	&X_1\\
X_2		&H_\text{bulk}\\
\end{pmatrix}
\end{equation}
where $H_\text{boundary}$ and $H_\text{bulk}$ are the topological boundary and bulk Hamiltonians respectively, $X_2$ is the coupling Hamiltonian for a particle hopping from the boundary to the bulk, and $X_1$ is the coupling Hamiltonian for the inverse hopping. Correspondingly, the effective boundary Hamiltonian reads
\begin{equation}
H_\text{eff} = H_\text{boundary}+X_1(E_\text{boundary}-H_\text{bulk})^{-1}X_2,
\label{eq:Heff}
\end{equation}
where $E_\text{boundary}$ is the eigenvalue corresponding to the topological boundary state and $(E_\text{boundary}-H_\text{bulk})^{-1}$ is the Green's function of the bulk. {The term $X_1(E_\text{boundary}-H_\text{bulk})^{-1}X_2$ is the affection of the bulk on topological boundaries (Fig.~\ref{sketch}). In the above, topological boundaries are the physical edges where topological boundary states reside.} Denoting the two topological boundary Hamiltonians by $h_A$ and $h_B$ respectively, the effective boundary Hamiltonian can be rewritten as
\begin{eqnarray}
H_\text{eff}=H_\text{boundary}+H_\text{perturb},
\label{eq:Heff2}
\end{eqnarray}
where
\begin{eqnarray}
&&H_\text{boundary}=
\begin{pmatrix}
h_A	&0\\
0	&h_B\\
\end{pmatrix}, \nonumber \\
&&H_\text{perturb}=
\begin{pmatrix}
\delta_A	&\Delta_\text{eff}\\
\Delta^\prime_\text{eff}	&\delta_B\\
\end{pmatrix},
\label{eq:H1H2}
\end{eqnarray}
with $\delta_A$ and $\delta_B$ ($\Delta_\text{eff}$ and $\Delta^\prime_\text{eff}$) being the diagonal (non-diagonal) terms of $X_1(E_\text{boundary}-H_\text{bulk})^{-1}X_2$. Note that the Green's function of the bulk $(E_\text{boundary}-H_\text{bulk})^{-1}$ encodes the information of the bulk on the topological boundaries. That is, the diagonal terms of $X_1(E_\text{boundary}-H_\text{bulk})^{-1}X_2$, i.e., $\delta_A$ and $\delta_B$, describe the effective on-site potentials of the topological boundaries, and the non-diagonal terms of $X_1(E_\text{boundary}-H_\text{bulk})^{-1}X_2$, i.e., $\Delta_\text{eff}$ and $\Delta^\prime_\text{eff}$, describe the effective couplings between the topological boundaries.

Before proceeding further, we introduce the notion of the non-normality of a matrix $M$. If the matrix $M$ is diagonalizable, i.e., $M = V \Lambda V^{-1}$ with $V$ nonsingular and $\Lambda$ diagonal, its non-normality is defined as
\begin{equation}
\kappa(V) = \sigma_\text{max}(V)/\sigma_\text{min}(V),
\end{equation}
where $\sigma_\text{max}(V)$ and $\sigma_\text{min}(V)$ are the largest and smallest singular values of $V$, respectively. Clearly, $\kappa(V)\geq1$. When $\kappa(V)=1$, the matrix $M$ is said to be normal; and when $\kappa(V)>1$, the matrix $M$ is said to be non-normal. Note that $\kappa(V)$ can be very large if $M$ is close to a non-diagonalizable point (dubbed as an exceptional point). As a result, if the matrix $M$ is non-diagonalizable, its non-normality diverges.

We now use the notion of the non-normality to analyze the spectrum and eigenstates of $H_{\text{eff}}$. It is known that $\Delta_\text{eff}$, $\Delta^\prime_\text{eff}$, $\delta_A$ and $\delta_B$ are small, so that $H_\text{perturb}$ can be seen as a perturbation to $H_\text{boundary}$. This means that to get the spectrum and eigenstates of $H_{\text{eff}}$, one can analyze the affection of $H_\text{perturb}$ on the spectrum and eigenstates of $H_\text{boundary}$. According to the Bauer-Fike theorem~\citep{bauer1960norms}, {when the product of the non-normality of $H_\text{boundary}$ and the matrix norm of the perturbation $H_\text{perturb}$ is larger than the minimal eigenvalue spacing of $H_\text{boundary}$~\citep{7SupMat}}, the pseudo-spectrum of $H_\text{boundary}$ can extend beyond its unperturbed spectrum, indicating that, in this case, the perturbation $H_\text{perturb}$ can induce significant changes to the spectrum of $H_\text{boundary}$. When this case occurs and according to the behavior of non-normal matrices~\citep{trefethen2020spectra}, the perturbation $H_\text{perturb}$ can also affect the non-normalities of the matrices $h_A$ and $h_B$, which means that the localization of the eigenstates of $H_\text{boundary}$ can be changed by the perturbation $H_\text{perturb}$.

Based on the above analysis, we summarize our mechanism as follows: the interplay between the non-normality of $H_\text{boundary}$ and the perturbation $H_\text{perturb}$ can affect the behavior of the higher-order boundary states and their corresponding eigenspectrum. In the following, we will demonstrate two phenomena induced by the mechanism: a new kind of phase transition and size-dependent higher-order topological boundary states.

\textit{\it \clr The phase transition.}---For convenience, we use the model of a 2D system with a pair of topological hinges {at the top and bottom edges (along $y$)} to demonstrate the phase transition. For this model, the characteristic equation is given by $f(\beta,E):=$Det$[H_\text{eff}^\text{2D}(\beta)-E]=0$, where $H_\text{eff}^\text{2D}(\beta)$ is the generalized Bloch Hamiltonian corresponding to the effective boundary Hamiltonian $H_\text{eff}^\text{2D}$ of this model, and the value of $E$ belongs to the OBC spectra of the topological boundary Hamiltonians of this model. $f(\beta,E)$ is an algebraic equation for $\beta$ with an even degree $2m$ in general, where $m$ denotes the rang of the hopping. The behavior of the spectrum and eigenstates of $H_\text{eff}^\text{2D}$ is determined by $\beta_m$ and $\beta_{m+1}$, where the solutions to the equation $f(\beta,E)$ are denoted by $\beta_j$ ($j=1,2,\cdots,2m$) with $|\beta_1|\leq\cdots\leq|\beta_{2m}|$. The relation between $\beta_m$ and $\beta_{m+1}$ can be classified as three cases: $\text{(i)}$ $|\beta_m|\neq|\beta_{m+1}|$ and $|\beta_m|\neq|\beta_{m+1}|^{-1}$; $\text{(ii)}$ $|\beta_m| = |\beta_{m+1}|$; and $\text{(iii)}$ $|\beta_m|=|\beta_{m+1}|^{-1}$. For case $\text{(i)}$ and the boundaries being finite size, $H_\text{eff}^\text{2D}$ gives a line-like spectrum and its eigenstates are skin localized. For the same case but under the thermodynamic limit, {i.e., the size of the system along the x direction being infinity}, $H_\text{eff}^\text{2D}$ gives a non-trivial point-gap and its eigenstates feature scale-free localization (SFL). For case $\text{(ii)}$ and regardless of the boundary sizes, $H_\text{eff}^\text{2D}$ gives a line-like spectrum and its eigenstates are skin localized. For case $\text{(iii)}$, there exists a critical boundary size. That is, when the boundary sizes are larger than the critical size, the topological boundary states feature SFL and the corresponding eigenenergies form a loop-like configuration. But when the boundary sizes are smaller than the critical size, the topological boundary states are skin localized and the corresponding eigenenergies form a line-like configuration. The results of the above three cases are described in detail in Section S2 of Supplementary Materials.

The above results indicate a new kind of phase transition: the transition between hybrid skin-topological states and scale-free topological boundary states. Note that hybrid skin-topological states have been researched widely, but the research on scale-free topological boundary states is severely lacking. {In particular, corner states with similar loop-like spectrum have usually also been considered as hybrid skin-topological states or higher-order non-Hermitian skin states\citep{okugawa2020second,kawabata_higher-order_2020,
palacios2021guided,kim2021disorder,li2023enhancement,ou_non-hermitian_2023,sun2024photonic}} In the following, we will give a criterion for the phase of scale-free topological boundary states.

To obtain the criterion, we derive the relation between the non-normality of the boundary Hamiltonian and $\beta_j$ with $j=m,m+1$. Using the topological origin of the NHSE, one can get that the non-normality $\kappa(V_\alpha)$ ($\alpha=1,2$) of the boundary Hamiltonian $h_\alpha$ grows exponentially with the boundary size $L$ (the two boundaries have the same size $L$), i.e., $\kappa(V_\alpha)\sim e^{c_\alpha{L}}$~\citep{nakai2024topological}, where $c_\alpha>0$ and the matrix $V_\alpha$ is the diagonalization matrix of $h_\alpha$. Then with the inverse localization length $-\ln|\beta_j|$ of the eigenstates of the boundary Hamiltonians, the relation between $\kappa(V_\alpha)$ and $\beta_j$ is
\begin{equation}
\kappa(V_\alpha)^{I_\alpha}\sim e^{I_\alpha{c_\alpha}L}\simeq|\beta_j|^L,
\end{equation}
where $I_\alpha$ is the sign of the spectral winding number
\begin{equation}
w_\alpha(E_r) = \int_0^{2\pi}\frac{dk}{2\pi i}\partial_k \log \det [h_\alpha(k)-E_r],
\end{equation}
with the reference energy $E_r$.
Using the above derived relation, the solutions to $f(\beta,E)$ and the effective couplings $\Delta_\text{eff}\simeq \delta^{L_\perp}$ and $\Delta^\prime_\text{eff}\simeq (\delta^\prime)^{L_\perp}$, we get the criteria: if the non-normalities of $h_A$ and $h_B$ satisfy the inequality~\citep{2SupMat}
\begin{equation}
e^{(I_Ac_A-I_Bc_B)} \leq |\delta\delta^\prime|^{L_\perp/L},
\end{equation}
the boundary states have {SFL} behavior and the corresponding spectrum features a point-gap. In the above, $\delta$ and $\delta^\prime$ are the functions of the parameters of the system's bulk, and $L_\perp$ is the size of the system along the direction perpendicular to the direction of the topological boundaries. The criteria we get indicates that the phase boundary between the point-gap and the line-like spectrum is dependent on the geometric aspect ratio $L_\perp/L$ of the system.

Note that some previous researches~\citep{okugawa2020second,kawabata_higher-order_2020,
palacios2021guided,kim2021disorder,li2023enhancement,ou_non-hermitian_2023,sun2024photonic,yang_percolation-induced_2024} have shown that the boundary spectra in 2D non-Hermitian systems under OBCs can have loop-like configurations. However, previous researches have neither provided an exact explanation for the origin of the loop-like spectra, nor shown the localization lengthes of the corresponding boundary states. Using the criteria we just got, one can see that these systems are in the phase of scale-free topological boundary states. More importantly, the behaviors of the spectra and the state localizations of these systems originate from our mechanism. Furthermore, based on our mechanism, we present similar results in 3D D-class second-order topological insulators, with the details provided in Sections S3, S4, and S5 of Supplementary Materials (see also references~\citep{zhang2018dynamical,lee2019anatomy,li_direct_2021,lei_classification_2022} therein).

\textit{\it \clr The size-dependent higher-order topological boundary states.}---Generally, the Hamiltonian matrices $h_A$ and $h_B$ in Eq.~\eqref{eq:Heff2} are Toeplitz, banded or semi-banded matrices. For a Toeplitz, banded or semi-banded matrix, as its size increases, its non-normality increases accordingly. Consequently, as the Hamiltonian matrix size ($h_A$ and $h_B$ have the same size) increases, the spectra of $h_A$ and $h_B$ are respectively close to their PBC spectra when the system is under the phase of scale-free topological boundary states~\citep{trefethen2020spectra}. That is, the OBC spectra of $h_A$ and $h_B$ are size-dependent when the system is in the phase of scale-free topological boundary states. Moreover, as the boundary size changes, such size-dependent spectra affect the gap close or reopen to assimilate the eigenenergies corresponding to the higher-order topological boundary states.

\begin{figure}[htb]
\includegraphics[width=1.0\linewidth]{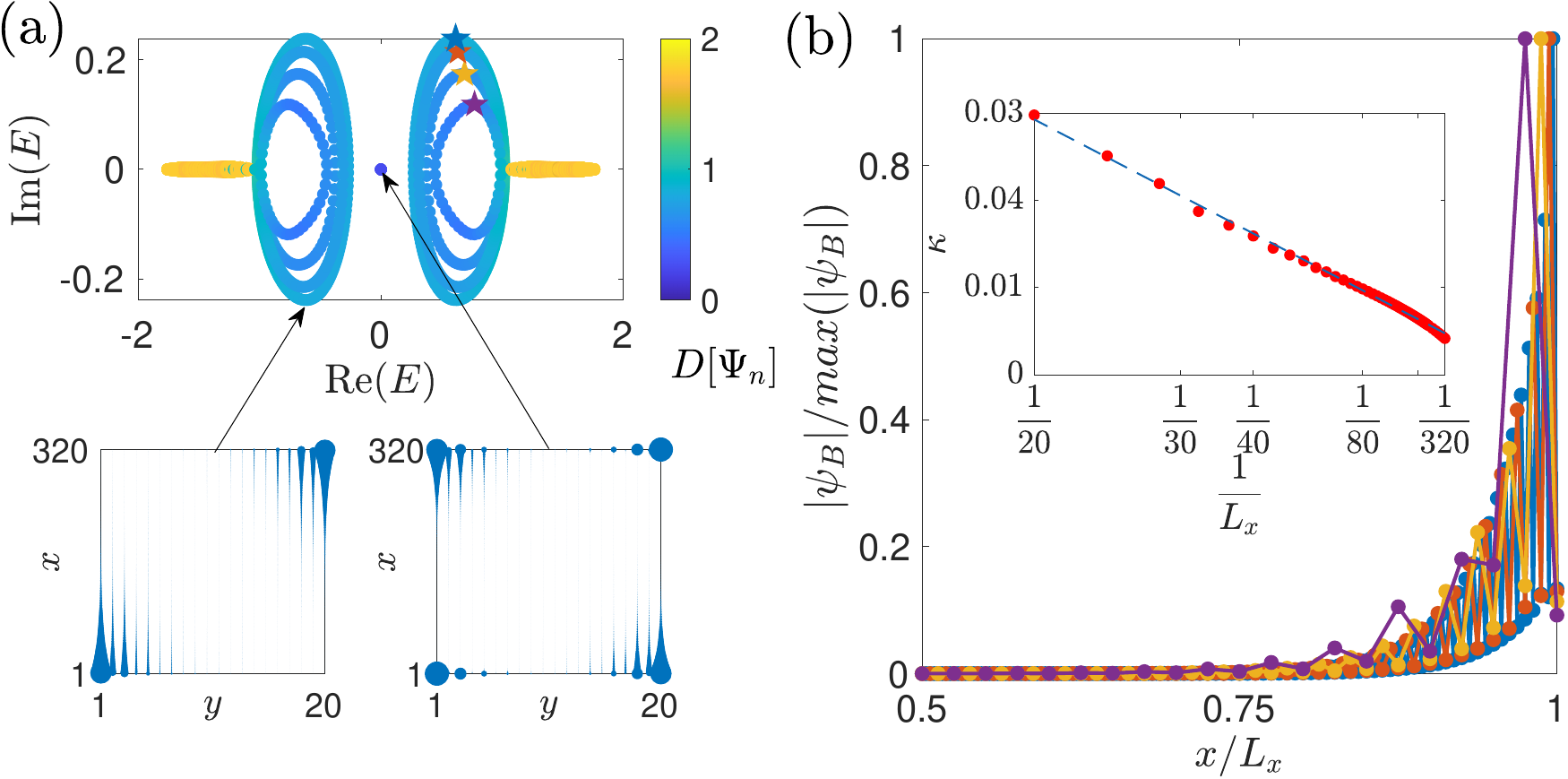}
\caption{{Different corner states in the 2D non-Hermitian BBH model.}
(a) The full-OBC spectra with $N_x = 20, 40, 80, 160$, and $N_y=10$. {The spectra are marked by different colors according to the fractal dimension $D_n[\Psi_n]=-\ln[\sum_{x,y}\vert\psi_{n}(x,y)\vert^4]/\ln\sqrt{L_x L_y}$, where $\Psi_n$ is an eigenstate and $\psi_{n}(x,y)$ is the wave amplitude of $\Psi_n$ at position $(x,y)$, which directly reveals the localization dimension of the state.}
The boundary states with $D[\Psi_n]\in [0,1]$ can be divided into two classes, respectively with loop spectra and zero eigenenergies. {The fractal dimension of these boundary states is less than 1 and close to 0, indicating their localization at the corners.}
As $N_x$ increases, the loop enlarges while the other features remain unchanged.
(b) Rescaled distribution of the boundary states with maximal imaginary energy (marked by the stars in (a)), showing a consistent profile across different $N_x$, along the top and bottom edges ($y=1$ and $y=L_y$) of the model.
The rescaled profile remains roughly unchanged for different $N_x$, indicating the SFL.
The inverse localization length $\kappa$ of the state with maximal imaginary energy is also found to linearly depend on $1/L_x$.
The other parameters are $t^\prime=0.25$, $t_y=t_x = 1$, $\gamma^+=-\gamma^- = 0.75$.}
\label{SFL}
\end{figure}

{In order to present the size-dependent feature more clearly, we consider a model where the bulk Hamiltonian is Hermitian while the topological boundary Hamiltonians are non-Hermitian.} Specifically, we consider the 2D BBH model \citep{benalcazar_quantized_2017,benalcazar_electric_2017} that supports corner localization induced by higher-order topology. The bulk Hamiltonian of the model reads
\begin{equation}
H(\bf{k}) = \begin{pmatrix}
0	&t^\prime + t_x e^{-ik_x}	&-t^\prime - t_y e^{ik_y}	&0\\
t^\prime + t_x e^{ik_x}	&0	&0	&t^\prime + t_y e^{ik_y}\\
-t^\prime - t_y e^{-ik_y}	&0	&0	&t^\prime + t_x e^{-ik_x}\\
0	&t^\prime + t_y e^{-ik_y}	&t^\prime + t_x e^{ik_x}	&0\\
\end{pmatrix},
\label{eq:bulkBBH}
\end{equation}
where $t_x$ and $t_y$ represent asymmetric inter-cell hopping parameters along $x$ and $y$ directions respectively, and $t^\prime$ is the amplitude of the intra-cell Hermitian hopping.
To make the boundary Hamiltonian of the BBH model non-normal, we introduce asymmetric non-Hermitian hopping amplitudes $\gamma^{\pm}$ to the top and bottom edges, respectively. As a result, the boundary Hamiltonian of the BBH model reads
\begin{equation}
H_\text{boundary}^\text{BBH}=\text{diag}[H_\text{SSH}^+,H_\text{SSH}^-] +\text{diag}[H_\text{nH}^+,H_\text{nH}^-],
\label{eq:edgeBBH}
\end{equation}
with
$H_\text{nH}^\pm = -i\gamma^\pm \cos k_x \sigma_y +i\gamma^\pm\sin k_x \sigma_x$. With Eqs.~\eqref{eq:Heff}, (\ref{eq:bulkBBH}) and (\ref{eq:edgeBBH}), one can get the effective boundary Hamiltonian~\citep{3SupMat},
{
\begin{equation}
\begin{aligned}
H_\text{eff}^\text{(nH-BBH)} =& \begin{pmatrix}
H^+_\text{nH-SSH}+\delta_+ 	&\Delta_+\\
\Delta_-		&H^-_\text{nH-SSH}+\delta_-\\
\end{pmatrix},\\
=&\left((t^\prime+t_x\cos k_x)\sigma_x+t_x\sin k_x\sigma_y\right)\tau_0\\
&+\left(i\gamma\sin k_x\sigma_x-i\gamma\cos k_x\sigma_y\right)\tau_z+\Delta\tau_x
\end{aligned}
\label{BBHeff}
\end{equation}}
with $H^\pm_{\rm nH-SSH}=(t^\prime+t_x\cos k_x+i\gamma^\pm\sin k_x)\sigma_x+(t_x\sin k_x-i\gamma^\pm \cos k_x)\sigma_y$
describing the top and bottom edges that form two identical 1D Su-Schrieffer-Heeger (SSH) chains \citep{su1979solitons}, $\delta_+ = \delta_-=0$, $\Delta = \Delta_+=\Delta_- = \frac{(t^\prime)^{N_y}}{(-1)^{N_y}(t_y)^{N_y-1}}\sigma_z$,
and $\sigma_{x,y,z}$ the Pauli matrices. Note that this model contains four sublattices within each unit cell,
and we use $N_{x(y)}=L_{x(y)}/2$ to denote the number of unit cells along $x(y)$ direction. {Furthermore, this effective Hamiltonian preserves PT-symmetry and CP-symmetry, i.e., $U_\text{PT}H_\text{eff}^\text{(nH-BBH)} U_\text{PT}^{-1}=(H_\text{eff}^\text{(nH-BBH)})^\ast$ and $U_\text{CP}H_\text{eff}^\text{(nH-BBH)} U_\text{CP}^{-1}=-(H_\text{eff}^\text{(nH-BBH)})^\ast$, where $U_\text{PT}=\sigma_x\tau_x$ and $U_\text{CP}=\sigma_y\tau_y$. When the above symmetries are broken, the hybrid skin-topological states of the model turn into scale-free topological boundary states~\citep{9SupMat}. Note that the SFL behavior is related to exceptional point phenomena, but not necessarily to a specific symmetry, such as PT-symmetry or CP-symmetry.}

{In this model, the original Hamiltonian $H(\textbf{k})$ preserves chiral symmetry, i.e., $\mathcal{C}H^\dagger \mathcal{C}^{-1} = -H(\textbf{k})$, as well as $C_4$ symmetry $\mathcal{R}_4H(k_x,k_y) \mathcal{R}_4^{-1} = H(k_y,-k_x)$ when $t_x =t_y$, where $\mathcal{C} = \tau_y\sigma_x$ is the chiral operator and gauge operator is
\begin{equation}
\mathcal{R}_4 = \begin{pmatrix}
0 &0 &-1	&0\\
1	&0	&0	&0\\
0	&0	&0	&1\\
0	&1	&0	&0\\
\end{pmatrix}.
\end{equation}
In the presence of non-normal boundary terms, one can obtain $\mathcal{C}(H_\text{boundary}^\text{BBH})^\dagger \mathcal{C}^{-1} = -H_\text{boundary}^\text{BBH}$ so that the chiral symmetry of the full model is still preserved. However, when $t_x =t_y$, only the bulk preserves the $C_4$ symmetry, and the full model preserves a $C_2$ symmetry. When $t_x\neq t_y$, the model preserves only a $C_2$ symmetry.}

Fig.~\ref{SFL}(a) demonstrates the OBC energy spectrum of the BBH model with non-Hermitian (top and bottom) edges, revealing two groups of corner states: the eigenenergies corresponding to the first group of corner states form an $N_x$-dependent loop-like spectrum, indicating that the first-order topological boundary states at opposite edges affect each other through the effective couplings $\Delta_+$, $\Delta_-$, and moreover the above affection induces the critical NHSE~\citep{li_criticalNHSE_2020,yokomizo_scaling_2021,qin_universal_2023} for the first-order topological boundary states at opposite edges; the eigenenergies corresponding to the second group of corner states remain at zero energy, suggesting their origin of higher-order topology, {as in the Hermitian BBH model.} {Note that the first group of corner states are the scale-free topological boundary states, and the second group of corner states are also termed zero-energy corner states.} Fig.~\ref{SFL}(b) demonstrates that the first group of corner states exhibit SFL, featuring roughly unchanged distribution profile rescaled with $L_x$, and the inverse localization length $\kappa$ (defined as $\psi(x,y)\sim e^{\kappa x}$) linearly depending on $L_x^{-1}$, while the second group of corner states possess almost the same localization length when changing the system's size (no shown). {Moreover, the scale-free topological boundary states are against the large boundary parameters~\citep{verma2024topological,zhang2024ptsymmetric} along the $x$-direction, whose details are shown in Section 7 of Supplement Materials.}

\begin{figure}[htb]
\includegraphics[width=1.0\linewidth]{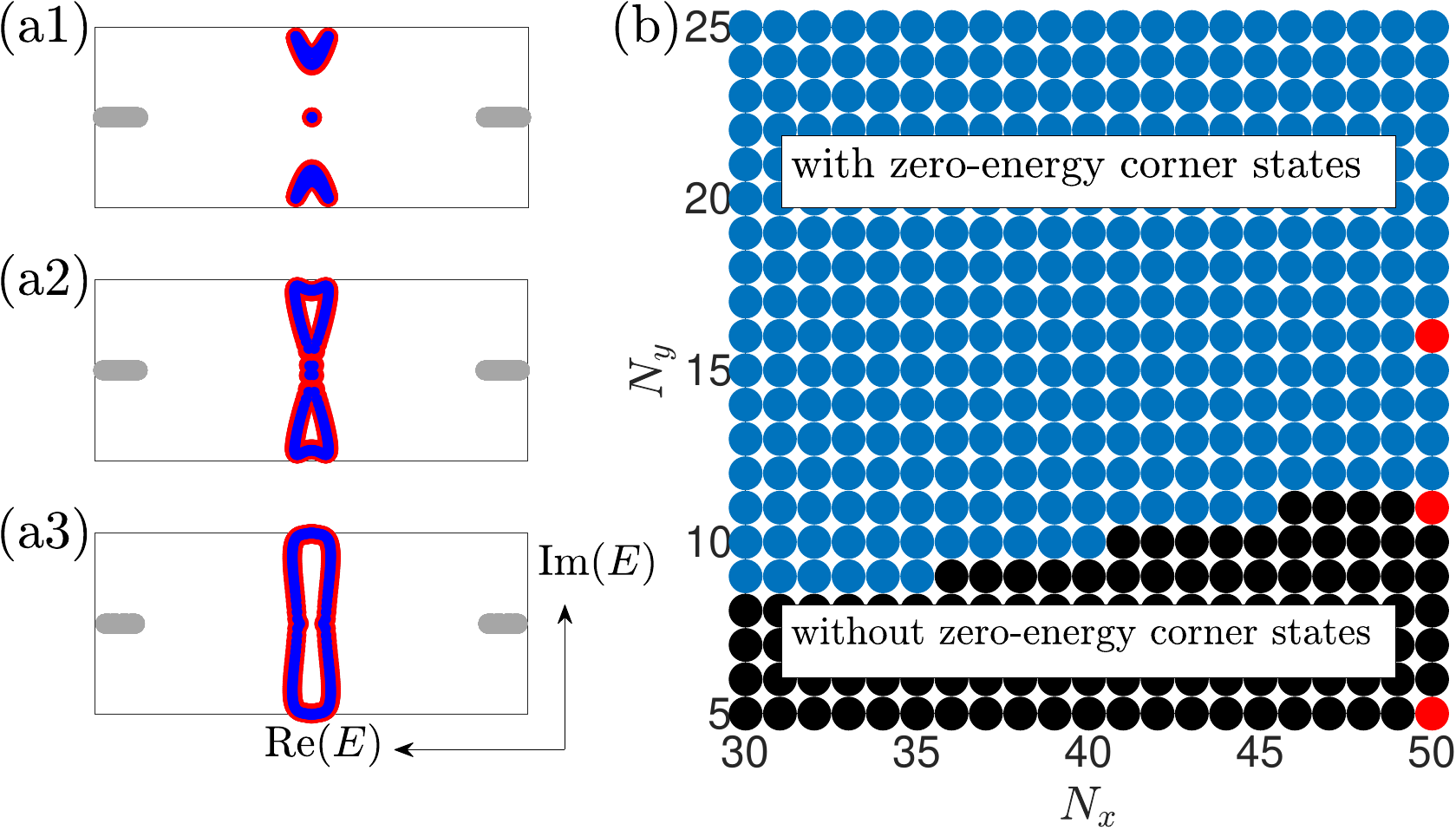}
\caption{Size-dependent {zero-energy} corner states in the 2D non-Hermitian BBH model.
(a1)-(a3) OBC spectra with $N_x=50$ and $N_y=16, 11, 5$, respectively,
of the model (red and gray) and the effective boundary Hamiltonian $H_{\rm eff}^{\rm (nH-BBH)}$ (blue).
(b) Phase diagram of the size-dependent {zero-energy} corner states. The other parameters are $t_x=0.4$, $\gamma^+=-\gamma^- = 0.6$, $t^\prime=0.3$, and $t_y=3.5$.}
\label{sizeDepend}
\end{figure}

Fig.~\ref{sizeDepend}(a) demonstrates that the two spectral loops corresponding to the {scale-free topological boundary states}, i.e., the first group of corner states, merge with each other when increasing $N_x$, closing the energy gap and eliminating the zero-energy corner states, i.e., the second group of corner states, when the gap reopens. It also demonstrates that the size-dependent spectral feature induced by the critical NHSE can alter the gap-closing condition, influencing the existence of the zero-energy corner states {(namely size-dependent zero-energy corner states)}. Fig.~\ref{sizeDepend}(b) demonstrates a phase diagram regarding the zero-energy corner states, versus the system's sizes $N_x$ and $N_y$. The behavior of the energy spectrum corresponding to the 1D topological states depends on the geometric ratio $N_x/N_y$ in the thermodynamic limit, which in turn influences the existence of {the zero-energy corner states}. This is consistent with our argument that the size-dependent energy spectrum, due to the interplay between the non-normalities of $H^{+(-)}_\text{nH-SSH}$ and the perturbations $\Delta_{+(-)}$, will directly affect the higher-order topological boundary states.

\textit{\it \clr Generalization.}---While our discussions in the preceding two sections are based on the models with 1D topological boundaries, the relevant results can be readily generalized to the models with ``unentangle" topological boundaries~\cite{jiang2023dimensional}. For a 3D system, its lowest-order topological states can be surface states. The Hamiltonians of its 2D topological boundaries can be ``unentangled" into separate sets of 1D chains $H_{1D}^1(k_1)\oplus H_{1D}^2(k_2)\oplus\cdots$, where $\{k_1,k_2,\cdots\}$ are the Bloch vectors along different directions.
As a simple example, we consider a 3D lattices system with a pair of 2D topological boundaries and the 2D topological boundary Hamiltonians are $H_\text{A(B)}(k_1,k_2) = H_{1D,\text{A(B)}}^1(k_1)\oplus H_{1D,\text{A(B)}}^2(k_2)$, where the two boundaries are denoted as A and B, respectively. When we open one boundary of these 2D topological boundaries to break the translation invariance corresponding to $k_1$, the boundary Hamiltonian reads
\begin{equation}
\begin{aligned}
&H_\text{A(B)}(x_1,k_2) = \sum_{x_1=1}^L\ket{x_1}\bra{x_1}\otimes H_{1D,\text{A(B)}}^2(k_2) \\
&\quad\quad+ \sum_{x_1=1}^{L-1}\bigg[\ket{x_1}\bra{x_1+1}\otimes U_{\text{A(B)}} + \ket{x_1+1}\bra{x_1}\otimes D_\text{A(B)}\bigg],
\end{aligned}
\end{equation}
where matrices $U_\text{A(B)}$ and $D_\text{A(B)}$ are derived by $H_{1D,\text{A(B)}}^1(k_1)$, $\{\ket{x_1}\}$ is the real space basis, and $L$ denotes the number of the lattice sites along the open boundary direction. Then, the effective boundary Hamiltonian of this system can be regarded as a pair of coupled 1D chains. Consequently, we can analyze the effective boundary Hamiltonian of this system with the similar procedures as those in the preceding two sections.

\textit{\it \clr Conclusion.}---We have uncovered a mechanism behind the behavior of higher-order boundary states in non-Hermitian systems: the interplay between the non-normality of $H_\text{boundary}$ and the perturbation $H_\text{perturb}$ can affect the behavior of the higher-order boundary states and their corresponding eigenspectrum. The mechanism stems from the sensitivity of the non-normal boundary Hamiltonian $H_\text{boundary}$ to the perturbation $H_\text{perturb}$, where the perturbation comes from the system itself, that is, the influence of the bulk on the topological boundaries. Leveraging this mechanism, we unveiled a novel type of phase transition: the transition between hybrid skin-topological states and scale-free topological boundary states;
the {latter} has been observed before but its origin remains obscure in previous literature \cite{okugawa2020second,kawabata_higher-order_2020,
palacios2021guided,kim2021disorder,li2023enhancement,ou_non-hermitian_2023,yang_percolation-induced_2024}.
Moreover, through this mechanism, we observed that scale-free topological boundary states exhibit spectra that vary with size.
Consequently, this variation influences the emergence of higher-order topological boundary states. Our findings pave the way for advancing the manipulation of higher-order boundary states and the topological characteristics of non-Hermitian systems.

In this study, we have focused on clean lattice models at the single-particle level to demonstrate the significant and often overlooked influence of intrinsic perturbations on non-normal subsystems - such as bulk perturbations affecting the boundary Hamiltonian in our model. Importantly, this mechanism arises fundamentally from the non-normality of the Hamiltonian matrices, suggesting its potential applicability to more complex scenarios. For instance, geometric defects within the bulk may exert a substantial impact on non-normal boundary states via the perturbation $H_{\rm perturb}$, even when located far from the physical boundaries~\cite{sun2021geometric,banerjee2024topological}. Beyond boundary-localization physics, this mechanism also holds promise for shedding light on other aspects of non-Hermitian systems, such as how perturbations from free particles might influence non-normal bound states in many-body systems~\cite{wang2023continuum,liu2024fate,qin2024occupation}.

\textit{\it \clr Acknowledgments.}---
The authors acknowledge support from the National Natural Science Foundation of China through Grant No. 12174224.

\bibliography{refs_BMBI}

\end{document}


\title{ Supplemental Material for
``Intrinsic-perturbation induced anomalous higher-order boundary states in non-Hermitian systems"}

\author{Hui-Qiang Liang}
\affiliation{Department of Physics, Shandong University, Jinan 250100, China}
\author{Zuxuan Ou}
\affiliation{Guangdong Provincial Key Laboratory of Quantum Metrology and Sensing $\&$ School of Physics and Astronomy, Sun Yat-Sen University (Zhuhai Campus), Zhuhai 519082, China}
\author{Linhu Li}\email{lilinhu@quantumsc.cn}
\affiliation{Quantum Science Center of Guangdong-Hong Kong-Macao Greater Bay Area (Guangdong), Shenzhen, China}
\author{Guo-Fu Xu}\email{xgf@sdu.edu.cn}
\affiliation{Department of Physics, Shandong University, Jinan 250100, China}

\date{\today}

\maketitle

\tableofcontents

\section*{S1. Derivation of the effective boundary Hamiltonian}
\subsection*{The $d-1$ dimensional boundary case}
Consider a general $d$-dimensional ($d\geq 2$) non-Hermitian lattice, under OBCs along $x$ direction with $L_x$ lattice sites and periodic boundary conditions (PBCs) along the other directions. Suppose this system has two $d-1$ dimensional topological boundaries at $x=1$ and $x=L_x$, both of which are coupled to environments. Without loss of generality, the Hamiltonian of this system can be written as
\begin{eqnarray}
\hat{H}=\hat{H}_\text{boundary}+\hat{H}_\text{bulk}+\hat{X}_1+\hat{X}_2,
\end{eqnarray}
where $\hat{H}_\text{boundary}$ and $\hat{H}_\text{bulk}$ are the boundary and bulk Hamiltonians respectively, $\hat{X}_1$ is the coupling Hamiltonian for a particle hopping from the bulk to the boundary, and $\hat{X}_2$ is the coupling Hamiltonian for the inverse hopping.

In the real space basis $\{\ket{s_1,\vec{k}},\ket{s_{L_x},\vec{k}},
\ket{s_2,\vec{k}},\cdots,\ket{s_{L_x-1},\vec{k}}\}$, the time-independent Schr\"odinger equation $\hat{H}\ket{\Psi}=E\ket{\Psi}$ can be written as
\begin{equation}
\begin{pmatrix}
H_\text{boundary}	&X_1\\
X_2	&H_\text{bulk}\\
\end{pmatrix}
\begin{pmatrix}
\Psi_\text{boundary}\\
\Psi_\text{bulk}\\
\end{pmatrix}
=E
\begin{pmatrix}
\Psi_\text{boundary}\\
\Psi_\text{bulk}\\
\end{pmatrix},
\label{eq:partition}
\end{equation}
where $s_x$ represents the $x$-th lattice site with $x\in\{1,2,\cdots,L_x\}$, and $\vec{k}$ is the quasi-momentum of the Bloch function in the PBC direction. In the above,
\begin{equation}
H_\text{boundary} = \begin{pmatrix}
h_A	&0\\
0	&h_B\\
\end{pmatrix}_{2\times2}
\label{eq:boundary}
\end{equation}
is the matrix representation of the operator $\hat{H}_\text{boundary}$ in the basis $\{\ket{s_1,\vec{k}},\ket{s_{L_x},\vec{k}}\}$, where $h_A$ and $h_B$ are the Bloch Hamiltonians for the pair of edges.
\begin{eqnarray}
&&H_\text{bulk} =\nonumber\\
&&\begin{pmatrix}
h_1^0		 &h_1^1	  	  &\cdots	&h_1^{r}	&0			&\cdots		&0\\
h_1^{-1}	 &h_2^{0} 	  &h_2^1	&\cdots		&h_2^{r}	&\ddots		&\vdots\\
\vdots		 &h_2^{-1}	  &h_3^0	&h_3^1		&\cdots		&\ddots		&0\\	
h_{1}^{-r}	 &\vdots  	  &h_3^{-1}	&\ddots		&\ddots		&\vdots		&h_{L_x-1-r}^{r}\\
0     	 	 &h_{2}^{-r}  &\vdots	&\ddots  	&h_{L_x-4}^0		& h_{L_x-4}^1&\vdots\\
\vdots		 &\ddots  	  &\ddots	&\cdots		&h_{L_x-4}^{-1}   &h_{L_x-3}^0	&h_{L_x-3}^1\\
0			 &\cdots	  &0		&h_{L_x-1-r}^{-r}&\cdots		&h_{L_x-3}^{-1}    &h_{L_x-2}^0 \\
\end{pmatrix}_{(L_x-2)\times (L_x-2)}
\label{eq:bulk}
\end{eqnarray}
is the matrix representation of the operator $\hat{H}_\text{bulk}$ in the basis $\{\ket{\vec{x}_2,\vec{k}},\cdots,\ket{\vec{x}_{L_x-2},\vec{k}}\}$ and it is a $r+1$-banded matrix, where $h_i^{0}$ is the Bloch Hamiltonian of the chains composing the bulk, and $h_i^{j}$ is the coupling between these chains, with $i\in\{1,2,\cdots,L_x-2-j\}$, $j\in\{-r,\cdots,r\}$ and $r$ being the maximal coupling range along $x$ direction.
\begin{equation}
\begin{aligned}
X_1 &= \begin{pmatrix}
c^1		&\cdots	&c^r	&0 &\cdots 	&0&\cdots&0\\
0&\cdots &0 &\cdots	&0 &c^r		&\cdots	&c^1\\
\end{pmatrix}_{2\times(L_x-2)},\\
X_2 &= \begin{pmatrix}
c^{-1}		&\cdots	&c^{-r}	&0 &\cdots 	&0&\cdots&0\\
0&\cdots &0 &\cdots	&0 &c^{-r}		&\cdots	&c^{-1}\\
\end{pmatrix}_{2\times(L_x-2)}^T
\end{aligned}
\label{x1x2}
\end{equation}
are the matrix representations of $\hat{X_1}$ and $\hat{X_2}$ respectively, where $T$ represents matrix transpose, the $1,j$ entry of $X_1$ and the $2,L_x-1-j$ entry of $X_2$ are respectively $c^j$ and $c^{-j}$, and all the unspecified entries of matrices $X_1$ and $X_2$ are zero, with $c^{j}$ and $c^{-j}$ being the couplings between the bulk and edges. According to Eq.~(\ref{eq:partition}), one can readily get
\begin{eqnarray}
&&X_2\Psi_\text{boundary}=(E-H_\text{bulk})\Psi_\text{bulk}, \label{eq:PsiBulk} \\
&&H_\text{boundary}\Psi_\text{boundary}+X_1\Psi_\text{bulk}=E\Psi_\text{boundary}. \label{eq:PsiEdge}
\end{eqnarray}

About the matrix $E-H_\text{bulk}$, we would like to make the following remark. As is well known, for a topological non-trivial system, the eigenenergies corresponding to the non-trivial topological boundary states lie within the bulk band gap protected by the bulk's symmetries. This ensures that the bulk states and boundary states do not overlap, and further prevents them from mixing or scattering. Specifically, as the eigenenergies corresponding to the boundary states and those corresponding to the bulk states approach each other, the boundary states begin to delocalize. And when these two kinds of eigenenergies mix with each other, the boundary states will merge into the bulk states. Thus, to avoid the above undesired situations, the eigenenergies corresponding to the boundary states must be far away from those corresponding to the bulk states. The above analysis indicates that $E$, as one of the eigenenergies corresponding to the boundary states, is sufficiently distant from the eigenvalues of $H_\text{bulk}$, which ensures the existence of the inverse of the matrix $E-H_\text{bulk}$.

Because the matrix $E-H_\text{bulk}$ is invertible, we can obtain from Eq.~(\ref{eq:PsiBulk}) the relation $\Psi_\text{bulk}=(E-H_\text{bulk})^{-1}X_2\Psi_\text{boundary}$, where $(E-H_\text{bulk})^{-1}$ is the corresponding inverse matrix. Then using the above relation and Eq.~(\ref{eq:PsiEdge}), we arrive at
\begin{eqnarray}
\left(H_\text{boundary}+X_1(E-H_\text{bulk})^{-1}X_2\right)\Psi_\text{boundary}
=E\Psi_\text{boundary},
\end{eqnarray}
which is the time-independent Schr\"odinger equation restricted to the system's boundaries. The above equation clearly shows that the effective boundary Hamiltonian corresponding to this system can be given by
\begin{equation}
H_\text{eff} = H_\text{boundary}+X_1(E-H_\text{bulk})^{-1}X_2.
\label{eq:Heff}
\end{equation}
We further derive the matrix form of the effective Hamiltonian in Eq.~(\ref{eq:Heff}). Denoting the $i,j$ entry of $(E-H_\text{bulk})^{-1}$ by $d_{i,j}$ and using Eq.~(\ref{x1x2}), the term $X_1(E-H_\text{bulk})^{-1}X_2$ can be expressed as
\begin{eqnarray}
&&X_1(E-H_\text{bulk})^{-1}X_2
=\begin{pmatrix}
\delta_A(E)		&\Delta_\text{eff}(E)\\
\Delta^\prime_\text{eff}(E)	&\delta_B(E)\\
\end{pmatrix}.
\label{eq10}
\end{eqnarray}
with
\begin{eqnarray}
&&\delta_A(E)=\sum_{i,j=1}^r d_{i,j}c^ic^{-j},\quad\delta_B(E)=\sum_{i,j=1}^r d_{L_x-(i+1),L_x-(j+1)}c^ic^{-j},  \nonumber \\
&&\Delta_\text{eff}(E)=\sum_{i,j=1}^r d_{i,L_x-(j+1)}c^ic^{-j},~\Delta^\prime_\text{eff}(E)=\sum_{i,j=1}^r d_{L_x-(j+1),i}c^jc^{-i}. \nonumber \\
\end{eqnarray}
Then according to Eqs.~(\ref{eq:boundary}) and (\ref{eq10}), the effective boundary Hamiltonian in Eq.~(\ref{eq:Heff}) turns into~\footnote[1]{Denote the eigenvalues and eigenstates of $H_\text{boundary}$ by $E^{(0)}_n$ and $\ket{\psi_n}$ respectively, the eigenvalues and eigenstates of $H_\text{bulk}$ by $\varepsilon_n$ and $\ket{\phi_m}$ respectively, the eigenenergies of $H$ by $E_n$. Then the entries of the matrix of the effective Hamiltonian $H_\text{eff}$ read $\bra{\Psi_{n^\prime}}H_\text{eff}\ket{\Psi_{n}} = \bra{\psi_{n^\prime}}H_\text{boundary}\ket{\psi_{n}}+\sum_{m}\frac{\bra{\psi_{n^\prime}}X_1\ket{\phi_m}\bra{\phi_m}X_2\ket{\psi_{n}}}{E_n-\varepsilon_m}$.}
\begin{eqnarray}
	H_\text{eff}=
	\begin{pmatrix}
h_A+\delta_A(E_A)	&\Delta_\text{eff}(E_A)\\
\Delta^\prime_\text{eff}(E_B)	&h_B+\delta_B(E_B)\\
\end{pmatrix}.
\label{Heff}
\end{eqnarray}

\subsection{The $d-n$ dimensional boundary case}
The above derivation can be readily generalized to the $d-n$ dimensional boundary case.
In the real space basis $\{\ket{\vec{b}_1,\vec{k}},\cdots,\ket{\vec{b}_{2^n},\vec{k}}; \ket{\vec{B}_1,\vec{k}},\cdots,\ket{\vec{B}_m,\vec{k}}\}$, where $\vec{b}_\mu$ represents the $\mu$-th boundary lattice site, $\vec{B}_{\nu}$ represents the $\nu$-th bulk lattice site and $\vec{k}$ represents the quasi-momentum of the Bloch function in the PBC direction, with the number of possible values for $\nu$ depending on the bulk's size. The time-independent Schr\"odinger equation $\hat{H}\ket{\Psi}=E\ket{\Psi}$ can be written as
\begin{equation}
\begin{pmatrix}
H_\text{boundary}	&X\\
K	&H_\text{bulk}\\
\end{pmatrix}
\begin{pmatrix}
\Psi_\text{boundary}\\
\Psi_\text{bulk}\\
\end{pmatrix}
=E
\begin{pmatrix}
\Psi_\text{boundary}\\
\Psi_\text{bulk}\\
\end{pmatrix}.
\end{equation}
In the above,
\begin{equation}
H_\text{boundary} = \text{Diagonal}\left[\{h_{b_1},h_{b_2},\cdots,h_{b_{2^n}}\}\right]_{2^n \times 2^n}
\end{equation}
is the matrix representation of the operator $\hat{H}_\text{boundary}$ in the basis $\{\ket{\vec{b}_1,\vec{k}},\cdots,\ket{\vec{b}_{2^n},\vec{k}}\}$ and it is a diagonal matrix, where $\{h_{b_1},h_{b_2},\cdots,h_{b_{2^n}}\}$ are the Bloch Hamiltonians for the edges.
\begin{eqnarray}
&&H_\text{bulk} =\nonumber\\
&&\begin{pmatrix}
h_1^0		 &h_1^1	  	  &\cdots	&h_1^{r}	&0			&\cdots		&0\\
h_1^{-1}	 &h_2^{0} 	  &h_2^1	&\cdots		&h_2^{r}	&\ddots		&\vdots\\
\vdots		 &h_2^{-1}	  &h_3^0	&h_3^1		&\cdots		&\ddots		&0\\	
h_{1}^{-r}	 &\vdots  	  &h_3^{-1}	&\ddots		&\ddots		&\vdots		&h_{m+1-r}^{r}\\
0     	 	 &h_{2}^{-r}  &\vdots	&\ddots  	&h_{m-2}^0		& h_{m-2}^1&\vdots\\
\vdots		 &\ddots  	  &\ddots	&\cdots		&h_{m-2}^{-1}   &h_{m-1}^0	&h_{m-1}^1\\
0			 &\cdots	  &0		&h_{m+1-r}^{-r}&\cdots		&h_{m-1}^{-1}    &h_{m}^0 \\
\end{pmatrix}_{m\times m}
\end{eqnarray}
is the matrix representation of the operator $\hat{H}_\text{bulk}$ in the basis $\{\ket{\vec{B}_1,\vec{k}},\cdots,\ket{\vec{B}_m,\vec{k}}\}$ and it is a $r+1$-banded matrix, where $h_i^0$ is the Bloch Hamiltonian of the chains composing the bulk, $h_i^{j}$ is the coupling between these chains and $r$ is the maximal coupling range in the bulk, with $i\in\{1,2,\cdots,m-j\}$ and $j\in\{-r,\cdots,r\}$. $X = (\chi_{\mu\nu})$ and $K = (\kappa_{\nu\mu})$ are the coupling matrices between the boundaries and bulk, where the entries $\chi_{\mu\nu} = \bra{\vec{b}_\mu,\vec{k}}\hat{H}\ket{\vec{B}_\nu,\vec{k}}$ and $\kappa_{\nu\mu} = \bra{\vec{B}_\nu,\vec{k}}\hat{H}\ket{\vec{b}_\mu,\vec{k}}$, with $\ket{\vec{b}_\mu,\vec{k}}\in\{\ket{\vec{b}_1,\vec{k}},\cdots,\ket{\vec{b}_{2^n},\vec{k}}\}$ and $\ket{\vec{B}_\nu,\vec{k}}\in\{\ket{\vec{B}_1,\vec{k}},\cdots,\ket{\vec{B}_m,\vec{k}}\}$.

With the similar derivations as those in the main text, the $d-n$ effective boundary Hamiltonian can be given by
\begin{equation}
H_\text{eff}(E) = H_\text{boundary}+X(E-H_\text{bulk})^{-1}K.
\end{equation}
where the nondiagonal elements of $X(E-H_\text{bulk})^{-1}K$ describe the effective couplings $\Delta$ in the system.

\section{S2. The eigenspectrum of a two-chain Hamiltonian and the localization of the corresponding eigenstates}
The effective boundary Hamiltonian for a system with a pair of 1D topological boundaries can be regarded as a  coupled two-chain system. In this section, we first calculate the condition for the point-gap spectrum and the localization of the corresponding eigenstates in a heuristic model, and then provide a general argument. Finally, we show the derivation of Eq.~(7) in the main text.

\subsection{The heuristic model}
We consider two weakly coupled Hatano-Nelson (HN) chains. The Hamiltonian matrix of this system reads
\begin{equation}
H = \begin{pmatrix}
h_A &0\\
0	&h_B\\
\end{pmatrix} + \begin{pmatrix}
D_{11}	&D_{12}\\
D_{21}	&D_{22}\\
\end{pmatrix},
\end{equation}
where second matrix is a perturbation to the first matrix, and the Hamiltonian matrices $h_A$ and $h_B$ of the open HN chains read
\begin{equation}
h_{A(B)} = \begin{pmatrix}
0	&t-g_{A(B)} 	&0	&\cdots\\
t+g_{A(B)} 	&0	&t-g_{A(B)} &\cdots\\
0	&t+g_{A(B)} &0&\cdots\\
\vdots	&\vdots	&\vdots	&\ddots\\
\end{pmatrix}.
\end{equation}
The Hamiltonian matrices $h_A$ and $h_B$ are non-normal when $g_{A(B)}\neq 0$. The diagonal entries $D_{11}$ and $D_{22}$ are the on-site potentials of the HN chains $A$ and $B$. Moreover, $D_{11}$ and $D_{22}$ do not affect the non-normality of $h_A$ and $h_B$. For convenience, we set $D_{11} = D_{22} = 0$. Then, the Hamiltonian matrix of this system can be rewritten as
\begin{equation}
H =\begin{pmatrix}
h_A	&D_{12}\\
D_{21}	&h_B\\
\end{pmatrix},
\label{eq:S19}
\end{equation}
which is same as the Hamiltonian matrix for the minimal model of the critical non-Hermitian skin effect (cNHSE).

By non-Bloch band theory~\citep{yao2018edge,yokomizo2019non}, the critical size for cNHSE has been given in Refs.~\citep{yokomizo_scaling_2021,qin_universal_2023}, and reads as
\begin{equation}
L_c+1 = \frac{2\ln|(t_L^At_R^B-t_R^At_L^B)^2f_{\infty}(E_\infty)\delta^2|}{\ln|t_L^Bt_R^A/(t_R^Bt_L^A)|},
\label{Lc}
\end{equation}
where
\begin{equation}
f_\infty(E_\infty)=\frac{\sqrt{E_\infty^2-4t_L^At_R^A}\sqrt{E_\infty^2-4t_L^Bt_R^B}}
{E_\infty^2(t_R^A-t_R^B)(t_L^A-t_L^B)+(t_L^At_R^B-t_R^At_L^B)^2}.
\end{equation}
Here, we take $t_L^A/t_R^A=R_A$, $t_L^B/t_R^B=R_B$ and $\sqrt{R_A}>\sqrt{R_B}$. To obtain the condition for the spectrum of the Hamiltonian matrix and featuring a point-gap, we substitute $k=\pi/2$ and $E_\infty=0$ into Eq.~\eqref{Lc}, and get
\begin{equation}
\begin{aligned}
&L_c+1 - \frac{2\ln|(t_L^At_R^B-t_R^At_L^B)^2f_{\infty}(E_\infty)\delta^2|}{\ln|t_L^Bt_R^A/(t_R^Bt_L^A)|} =0\\
\Rightarrow\qquad&\frac{R_B}{R_A}=\left((2\delta)^2(t_L^A t_R^B)\right)^{\frac{2}{L_c}},\\
\end{aligned}
\end{equation}
which means the point-gap phase will appear when $R_B/R_A\neq1$ and the system's size cross the critical size $L_c$.

The relation between $|\beta_3|$ and $|\beta_2|$ has been given in Refs.~\citep{yokomizo_scaling_2021,qin_universal_2023}, and reads as
\begin{equation}
\left\vert\frac{\beta_2}{\beta_3}\right\vert \simeq \left\vert(t_L^A t_R^B-t_L^B t_R^A)^2 f(E_\text{PBC})\delta^2 \right\vert^\frac{1}{L+1}
\label{Seq:b2b3}
\end{equation}
where $|\beta_3|>|\beta_2|$.
To obtain the phase boundary between NHSE and SFL, we substitute $|\beta_2/\beta_3|$ with $|\beta_{2c}/\beta_{3c}|$ in the above equation, where $|\beta_{2c}| = \sqrt{t_L^A/t_R^A}$ and $|\beta_{2c}| = \sqrt{t_L^B/t_R^B}$ correspond to the solutions to the characteristic equations of the HN chains, respectively. Eq.~\eqref{Seq:b2b3} becomes
\begin{equation}
\left\vert\frac{\beta_{2c}}{\beta_{3c}}\right\vert \simeq \left\vert(t_L^A t_R^B-t_L^B t_R^A)^2 f(E_\text{PBC})\delta^2 \right\vert^\frac{1}{L_c+1},
\label{Seq:b2b3c}
\end{equation}
where $L_c$ is the critical system size. By analyzing Eq.~\eqref{Seq:b2b3c},
one can see that the relation between $|\beta_{2c}|$ and $|\beta_{3c}|$ can be classified as three cases: $\text{(i)}$ $|\beta_{2c}|\neq|\beta_{3c}|$ and $|\beta_{2c}|\neq|\beta_{3c}|^{-1}$; $\text{(ii)}$ $|\beta_{2c}| = |\beta_{3c}|$; and $\text{(iii)}$ $|\beta_{2c}|=|\beta_{3c}|^{-1}$.

Case $\text{(i)}$: When the system is finite size and the system size is smaller than the critical size $L_c$, the system Hamiltonian gives a line-like spectrum and its eigenstates are skin localized.
When the system is finite size and the system size is larger than the critical size $L_c$, the spectrum of the system Hamiltonian is point gap and its eigenstates are skin localized.
For the same case but under the thermodynamic limit, the system Hamiltonian gives a non-trivial point-gap and its eigenstates feature SFL.

Case $\text{(ii)}$: The condition for the equality to hold is the system being under the thermodynamic limit. That is, regardless of the system size, the system Hamiltonian gives a line-like spectrum and its eigenstates are skin localized.

Case $\text{(iii)}$: When the system size is larger than the critical size, the eigenstates feature SFL and the corresponding eigenenergies form a loop-like configuration. But when the system size is smaller than the critical size, the eigenstates are skin localized and the corresponding eigenenergies form a line-like configuration.

\begin{figure}[h]
\includegraphics[width=1\linewidth]{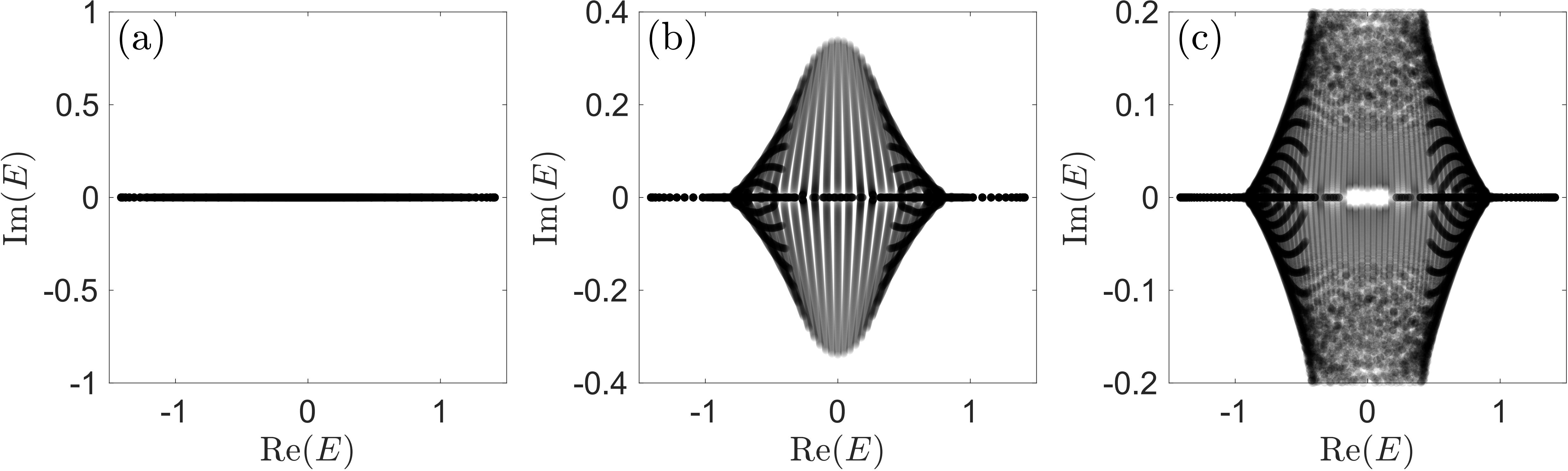}
\caption{The spectrum of Hamiltonian Eq.~\eqref{eq:S19}. The parameters are $t_L^A=1$, $t_R^A=0.5$, $R=[0.1,0.9]$, $\delta=0.01$. (a) shows the spectrum for $R_B/R_A=1$ case. (b) shows the spectrum for $R_B/R_A=R$ in $[0.1,0.9]$. The result of (c) is same as (b), but the system size in (c) is bigger than (b). One can infer that the point-gap are exist in thermodynamic limit. The other parameters are (a) $t_L^B = t_L^A R$, $t_R^B = t_R^A R$ and system size $L=40$; (b) $t_L^B =0.75$, $t_R^B = t_L^B R\frac{t_R^A}{t_L^A}$ and system size $L=40$; (c) (b) $t_L^B =0.75$, $t_R^B = t_L^B R\frac{t_R^A}{t_L^A}$ and system size $L=80$.}
\end{figure}

\subsection{The general case}
We start from a generic two-component ansatz Hamiltonian which reads
\begin{equation}
\mathcal{H}_g(z) =
\begin{pmatrix}
\mathcal{H}^{aa}(z)	&\mathcal{H}^{ab}(z)\\
\mathcal{H}^{ba}(z)	&\mathcal{H}^{bb}(z)\\
\end{pmatrix}
=\sum_{n=-n_-}^{n_+}
\begin{pmatrix}
h_n^{aa}	&h_n^{ab}\\
h_n^{ba}	&h_n^{bb}\\
\end{pmatrix}
z^n
\end{equation}
where $n_\pm\in \mathbb{Z}$, and $z=e^{ik}$. The components $\mathcal{H}^{aa}(z)$ and $\mathcal{H}^{bb}(z)$ have irreciprocal hoppings $h_n^{aa} \neq h_{-n}^{aa}$ and $h_n^{bb} \neq h_{-n}^{bb}$ for some $n$. In Ref.~\citep{qin_universal_2023}, the generalized Brillouin zone (GBZ) formalism of this Hamiltonian has been exhibited, and the corresponding result reads
\begin{equation}
\left\vert\frac{\beta_m}{\beta_{m+1}}\right\vert\simeq\left\vert-\frac{J(\beta_{i\in P_1},\beta_{j\in Q_1},E_\text{OBC})}{J(\beta_{i\in P_2},\beta_{j\in Q_2},E_\text{OBC})}\right\vert^{\frac{1}{L+1}}_{E_\text{OBC}=E_{\infty}},
\label{GBZ}
\end{equation}
where $P_1 = \{m+1,m+2,\cdots,2m\}$, $Q_1=\{1,2,\cdots,m\}$, $P_2 = \{m,m+2,m+3,\cdots,2m\}$, $Q_2=\{1,2,\cdots,m-2,m-1,m+1\}$, $m=n_++n_-$, and $L$ is the system size with $L\rightarrow\infty$.

In the thermodynamic limit $L\rightarrow\infty$, the right hand side of Eq.~\eqref{GBZ} tends to unity, giving rise to the standard GBZ solution $\vert\beta_m\vert = \vert\beta_{m+1}\vert$. The solution $\vert\beta_m\vert = \vert\beta_{m+1}\vert$ indicates that the spectrum of the system Hamiltonian is composed of the OBC spectra of the components $\mathcal{H}^{aa}(z)$ and $\mathcal{H}^{bb}(z)$. For the solution $\vert\beta_m\vert \neq \vert\beta_{m+1}\vert$, the spectrum of the system Hamiltonian is not composed of the OBC spectra of the components $\mathcal{H}^{aa}(z)$ and $\mathcal{H}^{bb}(z)$ in the thermodynamic limit. The couplings $\mathcal{H}^{ab}(z)$ and $\mathcal{H}^{ba}(z)$ can be regarded as a perturbation, and the spectrum of the system Hamiltonian is close to the composition of the PBC spectra of the components $\mathcal{H}^{aa}(z)$ and $\mathcal{H}^{bb}(z)$ in the thermodynamic limit. The reason is that when the size of a Toeplitz matrix is very large, the spectrum of the Toeplitz matrix is close to the generating function $f(z)=\sum_kt_kz^k$ (the PBC spectrum)~\citep{trefethen2020spectra}. The above argument suggests that these matrices have a critical size to distinguish the phase of OBC spectrum (spectra winding number being zero) and the phase of point gap spectrum. Base on this point, we can take the same analysis procedure for the solutions $\vert\beta_m\vert$ and $\vert\beta_{m+1}\vert$ as in the preceding subsection, where $\vert\beta_m\vert$ and $\vert\beta_{m+1}\vert$ are the the solutions to the characteristic equations of the component, respectively.

\subsection{The derivation of Eq.~(7) in the main text}
For a non-Hermitian system with non-zero winding PBC spectrum, the corresponding OBC system exhibits the NHSE and the NHSE origins from the non-normality $\kappa(V)\sim e^{cL}$ of the OBC system Hamiltonian~\citep{nakai2024topological}. Based on non-Bloch band theory, the skin modes have the inverse localization length $\ln|\beta_m|$. Therefore, we can find out
\begin{equation}
|\beta_m|^L \simeq \kappa(V)^I,
\label{kappa}
\end{equation}
where $I$ is the sign of the spectral winding number
\begin{equation}
w(E_r) = \int_0^{2\pi}\frac{dk}{2\pi i}\partial_k \log \det [H(k)-E_r],
\end{equation}
with the reference energy $E_r$. Using Eq.~\eqref{kappa} and Eq.~\eqref{GBZ}, one can obtain that
\begin{equation}
\left\vert\frac{\kappa(V_A)^{I_A}}{\kappa(V_B)^{I_B}}\right\vert\simeq\left\vert-\frac{J(\beta_{i\in P_1},\beta_{j\in Q_1},E_\text{OBC})}{J(\beta_{i\in P_2},\beta_{j\in Q_2},E_\text{OBC})}\right\vert_{E_\text{OBC}=E_{\infty}} = |\mathcal{J}(h_{n}^{aa},h_{n}^{bb},h_{nc}^{ab},h_{nc}^{ba},E_\text{OBC})|_{E_\text{OBC}=E_{\infty}},
\label{kAkB}
\end{equation}
where $\kappa(V_A)$ and $\kappa(V_B)$ are the non-normalities of the Hamiltonians of the components A and B, respectively; $I_A$ and $I_B$ are the the signs of the spectral winding numbers, $h_{nc}^{ab}$ and $h_{nc}^{ba}$ are the critical quantity of $h_{n}^{ab}$ and $h_{n}^{ba}$, respectively. The parameters $h_{nc}^{ab}$ and $h_{nc}^{ba}$ can be treated as perturbations. Expanding the function $|\mathcal{J}(h_n^{aa},h_n^{bb},h_{nc}^{ab},h_{nc}^{ba},E_\text{OBC})|_{E_\text{OBC}=E_{\infty}}$ to the leading order in terms of the perturbation max$\left\{|h_{nc}^{ab}h_{nc}^{ba}|\right\}$, Eq.~\eqref{kAkB} becomes
\begin{equation}
\left\vert\frac{\kappa(V_A)^{I_A}}{\kappa(V_B)^{I_B}}\right\vert\simeq \text{max}\left\{|h_{nc}^{ab}h_{nc}^{ba}|\right\} j(h_n^{aa},h_n^{bb},E_\text{OBC})_{E_\text{OBC}=E_{\infty}}.
\label{Seq:S30}
\end{equation}
Eq.~\eqref{Seq:S30} indicates the phase boundary between NHSE and SFL, and it suggests that the system exhibits NHSE and the eigenspecgtrum features line-like configuration when max$\left\{|h_{n}^{ab}h_{n}^{ba}|\right\}\leq$max$\left\{|h_{nc}^{ab}h_{nc}^{ba}|\right\}$. Therefore, the system is in the phase of NHSE when the following inequation is true,
\begin{equation}
\left\vert\frac{\kappa(V_A)^{I_A}}{\kappa(V_B)^{I_B}}\right\vert \geq \text{max}\left\{|h_n^{ab}h_n^{ba}|\right\} j(h_n^{aa},h_n^{bb},E_\text{OBC})_{E_\text{OBC}=E_{\infty}}>\text{max}\left\{|h_n^{ab}h_n^{ba}|\right\}.
\end{equation}
From the meaning of the above inequation, one can see that the following inequation
\begin{equation}
\left\vert\frac{\kappa(V_A)^{I_A}}{\kappa(V_B)^{I_B}}\right\vert \leq \text{max}\left\{|h_n^{ab}h_n^{ba}|\right\}
\label{eq:KaKb}
\end{equation}
means that the system Hamiltonian exhibits the point-gap spectrum and the corresponding eigenstates are SFL when the system parameters satisfy this inequation.

In the main text, we take a 2D system with a pair of topological hinges as an example. In this 2D system, $\text{max}\left\{|h_n^{ab}h_n^{ba}|\right\}$ is the effective coupling depending on the bulk's size, i.e., $\text{max}\left\{|h_n^{ab}h_n^{ba}|\right\} \Rightarrow |\Delta_\text{eff}\Delta_\text{eff}^\prime|\simeq |\delta\delta^\prime|^{L_\perp}$.
Here, $\delta$ and $\delta^\prime$ are the functions of the parameters of the system's bulk, and $L_\perp$ is the size of the system along the direction perpendicular to the direction of the topological boundaries. Using the relations $\kappa(V_A)^{I_A/L}\sim e^{I_A{c}_A}$ and $\kappa(V_B)^{I_B/L}\sim e^{I_B{c}_B}$, Eq.~\eqref{eq:KaKb} can be rewritten as
\begin{equation}
e^{(I_Ac_A-I_Bc_B)} \leq |\delta\delta^\prime|^{L_\perp/L},
\end{equation}
where $L$ is the size of the topological boundaries (the two boundaries have the same size), and the system parameters $c_A>0$ and $c_B>0$.

\section*{S3. Derivation of the effective boundary Hamiltonian of the BBH model}
\label{S1}
The Hamiltonian of the BBH model under $x$-PBCs and $y$-OBCs reads
\begin{equation}
H^\text{(BBH)}_{x-\text{PBC},y-\text{OBC}} = \begin{pmatrix}
h_+^{y=1}		&c_1	&0		&\cdots	&0		&0\\
c_1		&h_-^{y=2}	&c_2	&\cdots	&0		&0\\
0		&c_2	&h_+^{y=3}	&\cdots	&0		&0\\
\vdots	&\vdots	&\vdots	&\ddots	&\vdots	&\vdots\\
0		&0		&0		&\cdots	&h_+^{y=L_y-1}	&c_1\\
0		&0		&0		&\cdots	&c_1	&h_-^{y=L_y}\\
\end{pmatrix}_{L_y\times L_y},
\end{equation}
where $h_\pm = (t^\prime+t_x\cos k_x)\sigma_x+t_x\sin k_x\sigma_y$ are the Bloch Hamiltonians of the SSH chains, $c_1=-t^\prime\sigma_z$ and $c_2=-t_y\sigma_z$.
Note that this model contains four sublattices in a unit cell, and the degree of freedom of the $x$-direction sublattice is described by the Pauli's matrices $\sigma_{x,y,z}$, and the degree of freedom of the $y$-direction sublattice is denoted by the subscript $\pm$ of $h_\pm$.

Applying our method, we obtain the effective Hamiltonian
\begin{equation}
H_\text{eff} = H_\text{SSH-chain} + X(E-H_\text{bulk})^{-1}X^\dagger,
\end{equation}
where
\begin{equation}
H_\text{SSH-chain} = \begin{pmatrix}
h_+^{y=1}	&0\\
0	&h_-^{y=L_y}\\
\end{pmatrix},
\end{equation}
\begin{equation}
E-H_\text{bulk} = \begin{pmatrix}
E-h_-	&-c_2	&\cdots	&0\\
-c_2	&E-h_+	&\cdots	&0\\
\vdots	&\vdots	&\ddots	&\vdots	\\
0		&0		&\cdots	&E-h_+\\
\end{pmatrix}_{(L_y-2)\times(L_y-2)},
\end{equation}
and
\begin{equation}
X = \begin{pmatrix}
c_1	&0	&\cdots	&0	&0\\
0	&0	&\cdots	&0	&c_1\\
\end{pmatrix}_{2\times(L_y-2)}.
\end{equation}
Denoting by $d_{ij}$ the $i,j$ submatrix, which is a $2\times 2$ matrix, of the inverse of the matrix $E-H_\text{bulk}$, we can get
\begin{equation}
X(E-H_\text{bulk})^{-1}X^\dagger=\begin{pmatrix}
c_1 d_{1,1} c_1	&c_1 d_{1,L_y-2} c_1\\
c_1 d_{L_y-2,1} c_1	&c_1 d_{L_y-2,L_y-2} c_1\\
\end{pmatrix}.
\label{eq:dress_BBH}
\end{equation}
where
\begin{equation}
d_{ij} = \frac{\text{adj}(E-H_\text{bulk})_{ij}}{\text{Det}[E-H_\text{bulk}]}, \label{deij}
\end{equation}
with adj representing adjoint matrix and Det representing determinant. Note that all the submatrices of $E-H_\text{bulk}$, each of which is a $2\times 2$ matrix, commute with each other, and this is reason why one can use Eq.~(\ref{deij}) to carry out the calculation. To get $d_{ij}$, we first calculate the determinant Det$[E-H_\text{bulk}]$. Note that the value of $E$ is not fixed and it can take the values of $h_+$ and $h_-$, where $h_+$ and $h_-$ are the eigenvalues of $H_\text{SSH-chain}$. By using the Laplace expansion, one can readily get that the determinant Det$[E-H_\text{bulk}]$ reads
\begin{equation}
\text{Det}[h_+-H_\text{bulk}]=\text{Det}[h_--H_\text{bulk}] = (-c_2^2)^\frac{L_y-2}{2}.
\end{equation}
Accordingly, the four submatrices of $X(E-H_\text{bulk})^{-1}X^\dagger$ in Eq.~\eqref{eq:dress_BBH} read
\begin{equation}
\begin{aligned}
c_1 d_{1,1} c_1 &= c_1\frac{\text{adj}(h_+-H_\text{bulk})_{1,1}}{\text{Det}[h_+-H_\text{bulk}]}c_1 =
\begin{pmatrix}
0	&0\\
0	&0\\
\end{pmatrix},\\
c_1 d_{1,L_y-2} c_1 &= c_1\frac{\text{adj}(h_+-H_\text{bulk})_{1,L_y-2}}{\text{Det}[h_+-H_\text{bulk}]}c_1 = \frac{c_1^{N_y}}{(-1)^{N_y-1}c_2^{N_y-1}}\\
c_1 d_{L_y-2,1} c_1 &= c_1\frac{\text{adj}(h_--H_\text{bulk})_{L_y-2,1}}{\text{Det}[h_+-H_\text{bulk}]}c_1 = \frac{c_1^{N_y}}{(-1)^{N_y-1}c_2^{N_y-1}}\\
c_1 d_{L_y-2,L_y-2} c_1 &= c_1\frac{\text{adj}(h_--H_\text{bulk})_{L_y-2,L_y-2}}{\text{Det}[h_+-H_\text{bulk}]}c_1 = \begin{pmatrix}
0	&0\\
0	&0\\
\end{pmatrix}.\\
\end{aligned}
\end{equation}
The anti-diagonal elements, which describe the effective couplings $\Delta$, show a dependence on the size of the system:
\begin{equation}
\Delta_\text{eff}(E=h_+)=\Delta_\text{eff}(E=h_-) = c_1d_{1,L_y-2}c_1 = c_1d_{L_y-2,1}c_1 = \frac{c_1^{N_y}}{(-1)^{N_y-1}c_2^{N_y-1}}.
\end{equation}

\begin{figure}[hbt]
\includegraphics[width=0.7\linewidth]{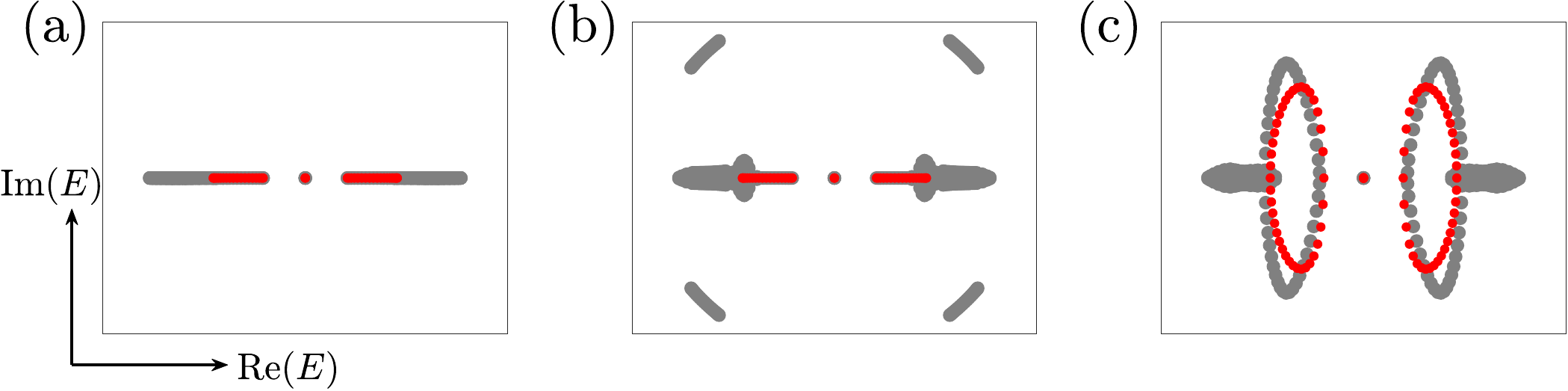}
\caption{The spectra of the effective Hamiltonian and system Hamiltonian. (a) The spectra of the Hermitian Hamiltonian and its boundary effective Hamiltonian. (b) The spectra of the non-Hermitian Hamiltonian with the same skin for each chain and its boundary effective Hamiltonian. (c) The spectra of the Hermitian bulk's Hamiltonian with the opposite skin for the boundaries and its boundary effective Hamiltonian. The parameters of $x$-axis SSH chain are $t^\prime = 0.25$, (a) $t_x = \sqrt{1.75\times 0.25}$, (b) $t_x = 1$, $\gamma = 0.75$, and (c) $t_x = 1$, $\gamma = 0.75$. The other parameters are $N_x = 20$, $N_y=10$, and $t_y = 1$.}
\label{Hedge}
\end{figure}

In Fig.~\ref{Hedge}, we display the comparison between the spectrum of the BBH model and the spectrum of the boundary effective Hamiltonian for different cases. In those cases, we choose the relation $E_\pm = h_\pm$, where the diagonal terms disappear in $X(E-H_\text{bulk})^{-1}X^\dagger$.
Fig.~\ref{Hedge}(a) and (b) show that the spectra of the boundary effective Hamiltonian are consistent with the eigenenergies of the topological boundary states of the system. In Fig.~\ref{Hedge}(b), the Hamiltonian describing the chains in the bulk reads
\begin{equation}
h_\pm^\prime = h_\pm+\gamma \cos k \sigma_y +i\gamma\sin k \sigma_x.
\end{equation}
The above Hamiltonian prevents cNHSE from occurring because opposite boundaries have the same skin, while the relation $E_\pm = h_\pm$ remains valid even as the system's size changes. In Fig.~\ref{Hedge}(c), the system is coupled to the environment through the boundaries. This figure reveals that the eigenenergies of the effective Hamiltonian are not consistent with the topological boundary states, and the reason is that $E_\pm$ are too close to the eigenenergies of $H_\text{bulk}$.

\section*{S4. Point-gap persisting under OBCs for 3D D-class HOTPs}
\begin{figure*}[ht]
\includegraphics[width=1.0\linewidth]{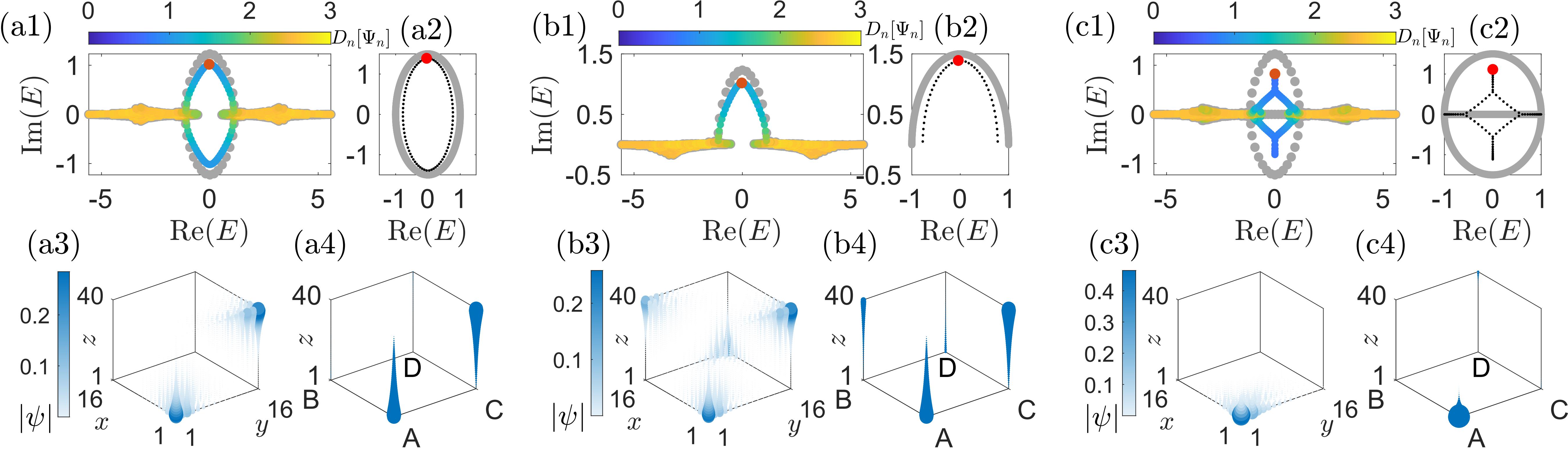}
\caption{
(a1/b1/c1) The $x,y$-OBC spectra of the 3D Hamiltonian in Eq.~(\ref{eq:Dclass})
with the non-Hermitian boundaries described by Table \ref{Table1}. The spectra marked by different colors according to the fractal dimension $D_n[\Psi_n]=-\ln(\sum_{x,y,z}|\psi_n(x,y,z)|^4)/\ln\sqrt[3]{L_xL_yL_z}$. (a2/b2/c2) The PBC (gray dots) and OBC (black dots) spectra of $H_\text{eff}^\text{(nH-D)}$ for $k_z\in[-\pi/2,\pi/2]$.
(a3/b3/c3) The distributions of the hinge states. (a4/b4/c4) The distributions of the eigenstates in the effective model. Both of the states are marked as red dots on the spectra in (a1/b1/c1) and (a2/b2/c3), respectively. The parameters are $N_x=N_y=8$, $m_1=-1.5$, $m_2=-0.5$, $\gamma=0.75$ and $N_z = 40$.}
\label{DclassSkin}
\end{figure*}
For 3D system, we show the result for second-order topological insulator~\citep{lei_classification_2022} in Fig.~\ref{DclassSkin}. The bulk Hamiltonian of this model reads
\begin{equation}
H^{(\text{D})}(\textbf{k}) = H_1(\textbf{k})+H_2(\textbf{k}_{1,\parallel}),
\label{eq:Dclass}
\end{equation}
where $H_1(\textbf{k}) = M_1(\textbf{k})\sigma_x\tau_0+h_1(\textbf{k})\sigma_y\tau_0$
and $H_2(\textbf{k}_{1,\parallel}) =M_2(\textbf{k}_{1,\parallel})\sigma_z\tau_x+h_2(\textbf{k}_{1,\parallel})\sigma_z\tau_y
+h_3(\textbf{k}_{1,\parallel})\sigma_z\tau_z$, with $M_1 = m_1+3-\cos k_x-\cos k_y-\cos k_z$, $M_2=m_2+2-\cos k_y-\cos k_z$, $h_1=\sin k_x$, $h_2=\sin k_y$ and $h_3=\sin k_z$. Here, $\sigma_{\beta=x,y,z}$ and $\tau_{\beta=x,y,z}$ are two sets of Pauli matrices, and $\tau_0$ is a two-by-two identity matrix. In a region of $k_z$ centered at $k_z=0$, this model possesses topological hinge states with different group velocities (see Table.~\ref{Table1}). Asymmetric non-Hermitian hopping $i 2\gamma_{\pm}\cos{k_z}$ is introduced to the four hinges, with $\gamma_+=-\gamma_-=\gamma$ inducing non-Hermitian pumping toward different directions for different hinges (see Table.~\ref{Table1}).
By applying our method to this model (see Supplement Materials S5), we obtain an effective boundary Hamiltonian
\begin{equation}
H_\text{eff}^\text{(nH-D)} = \begin{pmatrix}
h_{(1_x,1_y)}	&\Delta			&\delta		&0\\
\Delta			&h_{(N_x,1_y)}	&0			&\delta\\
\delta			&0				&h_{(1_x,N_y)}	&\Delta	\\
0				&\delta			&\Delta			&h_{(N_x,N_y)}\\
\end{pmatrix},
\label{eq:DclassHeff}
\end{equation}
where the subscripts of $h_{(x,y)}$ indicate the locations of different hinges, and $\Delta=\exp(-0.8804N_x-1.1345)$ and $\delta=\exp(-0.6991N_y-0.5378)$ describe the effective couplings

\begin{table}[htb]
\begin{center}
\begin{tabular}{c|c|c|c|c|c}
\hline
\multirow{2}{*}{}&Hinges & $(1_x,1_y)_A$ & $(N_x,1_y)_B$ & $(1_x,N_y)_C$ & $(N_x,N_y)_D$ \\
\cline{2-6}
&group velocity & $+$ & $-$ & $-$ & $+$\\
\hline
\hline
\multirow{2}{*}{Case I}&non-Hermiticity & $i\gamma_+\cos{k_z}$ & $i\gamma_-\cos{k_z}$ & $i\gamma_+\cos{k_z}$ & $i\gamma_-\cos{k_z}$ \\
\cline{2-6}
&nH-pumping & \blue{$+$} & \blue{$+$} & \red{$-$} & \red{$-$}\\
\hline
\multirow{2}{*}{Case II}&non-Hermiticity & $i\gamma_+\cos{k_z}$ & $i\gamma_+\cos{k_z}$ & $i\gamma_+\cos{k_z}$ & $i\gamma_+\cos{k_z}$ \\
\cline{2-6}
&nH-pumping & \blue{$+$} & \red{$-$} & \red{$-$} & \blue{$+$}\\
\hline
\multirow{2}{*}{Case III}&non-Hermiticity & $i\gamma_+\cos{k_z}$ & 0 & 0 & $i\gamma_-\cos{k_z}$ \\
\cline{2-6}
&nH-pumping & \blue{$+$} & $\setminus$ & $\setminus$ & \red{$-$}\\
\hline
\end{tabular}
\end{center}
\caption{
The signs of the group velocities $\partial E/\partial k_z$ of the hinge states and the signs of the non-Hermitian pumping (nH-pumping) for different non-Hermitian terms $i 2\gamma_{\pm}\cos{k_z}$.
The nH-pumping toward $z=1$ ($z=L_z$) is denoted by ``$+ ~($-$)$".
The three cases correspond to panels (a), (b), and (c) in Fig.~\ref{DclassSkin}, respectively.
{Such non-Hermitian terms break the particle-hole symmetry of the full system in cases II and III, and preserve the particle-hole symmetry in case I. The rotational symmetry does not break in cases I and II, and break in case III ($C_2\rightarrow C_1$).}}
\label{Table1}
\end{table}

To show the different behaviors of the hinge states, we consider the three cases in Table.~\ref{Table1}. Figures~\ref{DclassSkin}(a) and (b) show the results for cases I and II, respectively. From them, one can see that both of the corner states are SFL, and the corresponding hinge states acquire positive and negative imaginary energies respectively, forming a point-gap or semi-point-gap (i.e., boundary spectrum forming half loop) under full-OBCs. Figure~\ref{DclassSkin}(c) shows the result for case III. From it, one can see that the boundary spectrum transitions from $\theta$-like point-gap to diamond-like point-gap when changing from $z$-PBCs to $z$-OBCs. Moreover, one can also see that in the diamond-like point-gap phase, the boundary spectral loop is partially closed into some lines and the corner states are accumulated by the NHSE. Based on the setting of case III and Eq.~(\ref{eq:DclassHeff}), one can get that the perturbation of bulk between the pair of diagonal hinges is absent, avoiding critical NHSE and further causing the corner states to be accumulated by the NHSE. Additionally, the positive/negative imaginary energies indicate that the non-Hermitian pumping aligns/conflicts with the group velocities in the hinges. From Fig.~\ref{DclassSkin}, one can see that for all the three cases, the boundary spectral features and the distributions of the hinge or corner states are consistent with the effective Hamiltonian with the 1D momentum $k_z\in[-\pi/2,\pi/2]$, see S5.

\section*{S5. Derivation of the effective Hamiltonian of 3D D-class HOTPs}
\label{S5}
We consider a Hamiltonian with second-order topology belonging to the D class~\citep{lei_classification_2022},
\begin{equation}
H^{(\text{D})}(\textbf{k}) = H_1(\textbf{k})+H_2(\textbf{k}_{1,\parallel}),
\label{Seq:Dclass}
\end{equation}
where
\begin{equation}
\begin{aligned}
H_1(\textbf{k}) &= M_1(\textbf{k})\sigma_x\tau_0+h_1(\textbf{k})\sigma_y\tau_0,\\
H_2(\textbf{k}_{1,\parallel}) &=M_2(\textbf{k}_{1,\parallel})\sigma_z\tau_x+h_2(\textbf{k}_{1,\parallel})\sigma_z\tau_y+h_3(\textbf{k}_{1,\parallel})\sigma_z\tau_z,
\end{aligned}
\end{equation}
with $M_1 = m_1+3-\cos k_x-\cos k_y-\cos k_z$, $M_2=m_2+2-\cos k_y-\cos k_z$, $h_1=\sin k_x$, $h_2=\sin k_y$ and $h_3=\sin k_z$. Here, $\sigma_{\beta=x,y,z}$ and $\tau_{\beta=x,y,z}$ are two sets of Pauli matrices, and $\tau_0$ is a two-by-two identity matrix. According to the topological characterization of nested band-inversion surface (BIS)~\citep{zhang2018dynamical,li_direct_2021,lei_classification_2022}, this model supports second-order gapless hinge states under $(x,y)$-OBCs, crossing at zero energy when $k_z=0$, in the parameter regions of $-2<m_1<m_2<0$ and $-2<m_2<m_1+2<0$.
Figure~2 in the main text provides an example of the energy spectrum and hinge states of this model with $m_1=-1.5$ and $m_2=-0.5$.

To study the second-order hinge states of this model, consider the Hamiltonian under $(x,y)$-OBCs,
\begin{eqnarray}
H^{(\text{D})}_{(x,y)-\text{OBCs}}(\textbf{k}_z) &=& \sum_{n_x,n_y}\left(\tilde{M}_1\sigma_x\tau_0
+\tilde{M}_2\sigma_z\tau_x+h_3\sigma_z\tau_z\right)\hat{a}_{n_x,n_y}^\dagger \hat{a}_{n_x,n_y}
+\frac{-\sigma_x\tau_0+i\sigma_y\tau_0}{2}\hat{a}^\dagger_{x+1,y}\hat{a}_{n_x,n_y} \nonumber \\
&&+\frac{-\sigma_x\tau_0-\sigma_z\tau_x+i\sigma_z\tau_y}{2}
\hat{a}^\dagger_{n_x,n_y+1}\hat{a}_{n_x,n_y}+\text{h.c.},
\label{eq:DclassOBC}
\end{eqnarray}
where $\tilde{M}_1 = m_1+3-\cos k_z$, $\tilde{M}_2=m_2+2-\cos k_z$, $h_3=\sin k_z$ and $\hat{a}_{n_x,n_y}^\dag=(\hat{A}_{n_x,n_y}^\dagger,\hat{B}_{n_x,n_y}^\dagger,
\hat{C}_{n_x,n_y}^\dagger,\hat{D}_{n_x,n_y}^\dagger)^{T}$, with $\hat{a}_{n_x,n_y}^\dag$ the creation operator at $(n_x,n_y)$ site in different sublattices of a unit cell.
The two sets of Pauli's matrices $\sigma$ and $\tau$ act on the space of the sublattcies, labeled as A, B, C and D;
and the system is chosen to have $N_x\times N_y$ unit cells.
The hinges are localized at sites $\{(1,1)_A,(N_x,1)_B,(1,N_y)_C,(N_x,N_y)_D\}$, where the above subscripts denote the sublattices.

\begin{figure}[h]
\includegraphics[width=0.6\linewidth]{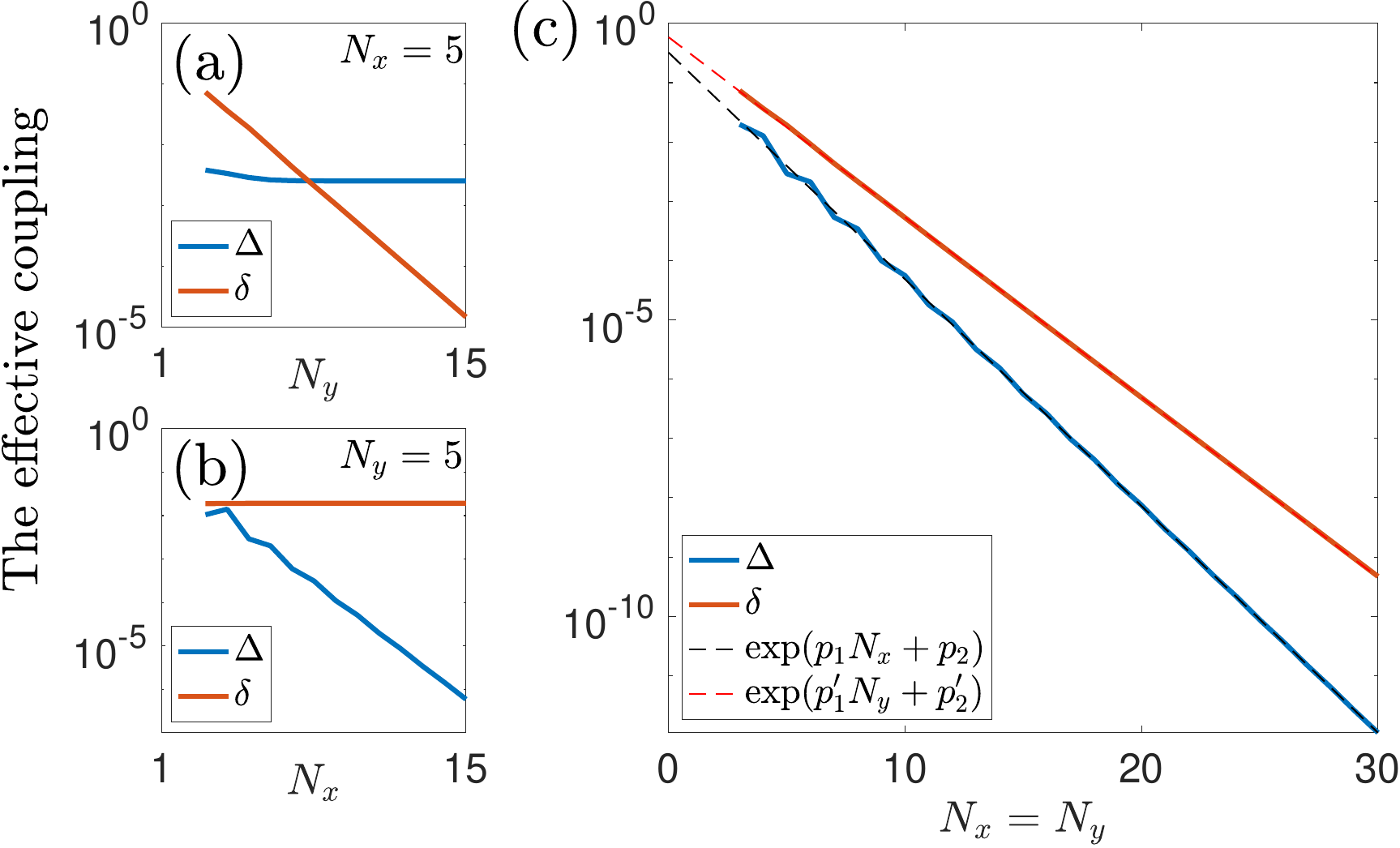}
\caption{The effective coupling for the hinge Hamiltonian in Eq.~(\ref{eq:hinge}), i.e. $\Delta$ and $\delta$, as a function of (a) $N_y$ with $N_x=5$ and (b) $N_x$ with $N_y=5$, respectively.
(c) $\Delta$ as a function of $N_x$ and $\delta$ as a function of $N_y$, with $N_x=N_y$. The solid lines represent the numerical results, and the dashed lines represent the fitting results of $\Delta = \exp{(p_1N_x+p_2)}$ and $\delta = \exp{(p_1^\prime N_y+p_2^\prime)}$, with $p_1 =-0.8804$, $p_2=-1.1345$, $p_1^\prime=-0.6991$, and $p_2^\prime=-0.5378$. The other parameters are $m_1=-1.5$ and $m_2=-0.5$.}
\label{Ceff}
\end{figure}

As discussed in the main text, the effective boundary Hamiltonian reads
\begin{eqnarray}
H_\text{eff}(E) = H_\text{boundary}+X(E-H_\text{bulk})^{-1}X^\dagger, \label{eq:Heff_formal_SM}
\end{eqnarray}
where $H_\text{boundary}$ corresponds to the term $h_3\sigma_z\tau_z$ in Eq.~\eqref{eq:DclassOBC}. We then numerically solve Eq.~\eqref{eq:Heff_formal_SM} under the condition of choosing the eigenenergies of the hinge states at $k_z=0$ which are furthest from the bulk bands. That is, we numerically solve the equation under the condition of $E=0$. As a result, we obtain the effective hinge Hamiltonian,
\begin{equation}
\begin{aligned}
H_\text{eff}^\text{(D)}=
\begin{pmatrix}
\sin{k_z}	&0	&0	&0\\
0 &-\sin{k_z}	&0	&0\\
0	&0	&-\sin{k_z}	&0\\
0	&0	&0	&\sin{k_z}\\
\end{pmatrix}
+
\begin{pmatrix}
0	&\Delta	&\delta	&0\\
\Delta &0	&0	&\delta\\
\delta	&0	&0	&\Delta\\
0	&\delta	&\Delta	&0\\
\end{pmatrix},
\end{aligned}
\label{eq:hinge}
\end{equation}
with the effective couplings $\Delta$ and $\delta$. Note that for the latter matrix in Eq.~(\ref{eq:hinge}), the nonzero entries of it are $\Delta$ or $\delta$ because of the 2D reflection symmetry of the system, i.e.,
\begin{equation}
H^{({\rm D})}(k_x,k_y,k_z)=[H^{({\rm D})}(-k_x,-k_y,k_z)]^*,
\end{equation}
with $*$ representing complex conjugate. The other entries of the matrix take the value of zero because (i) the hopping along the diagonal direction in the bulk do not exist, resulting in the anti-diagonal entries being zero, and (ii) the bulk only connects different boundaries without adding any onsite term, resulting in the diagonal entries being zero. The numerical and fitting results of the effective couplings are illustrated in Fig.~\ref{Ceff}. It is seen in panels (a) and (b) that $\Delta$ weakly depends only on $N_x$, and $\delta$ weakly depends only on $N_y$.
Therefore, we take the fitting functions of $\Delta$ and $\delta$ as $\Delta = \exp{(p_1N_x+p_2)}$ and $\delta = \exp{(p_1^\prime N_y+p_2^\prime)}$, as shown in Fig.~\ref{Ceff}(c). By further considering the non-Hermiticity of the hinges, we have
\begin{equation}
\begin{aligned}
H_\text{eff}^\text{(nH-D)}=
&H_\text{eff}^\text{(D)}+H_\text{nH}\\
=&
\begin{pmatrix}
\sin{k_z}	&\Delta	&\delta	&0\\
\Delta &-\sin{k_z}	&0	&\delta\\
\delta	&0	&-\sin{k_z}	&\Delta\\
0	&\delta	&\Delta	&\sin{k_z}\\
\end{pmatrix}
+
\begin{pmatrix}
2i\gamma_A\cos{k_z}	&0	&0	&0\\
0	&2i\gamma_B\cos{k_z}	&0&0\\
0&0&2i\gamma_C\cos{k_z}&0\\
0&0&0&2i\gamma_D\cos{k_z}\\
\end{pmatrix}.
\end{aligned}
\label{eq:hinge1}
\end{equation}

\begin{figure}[hbt]
\includegraphics[width=0.675\linewidth]{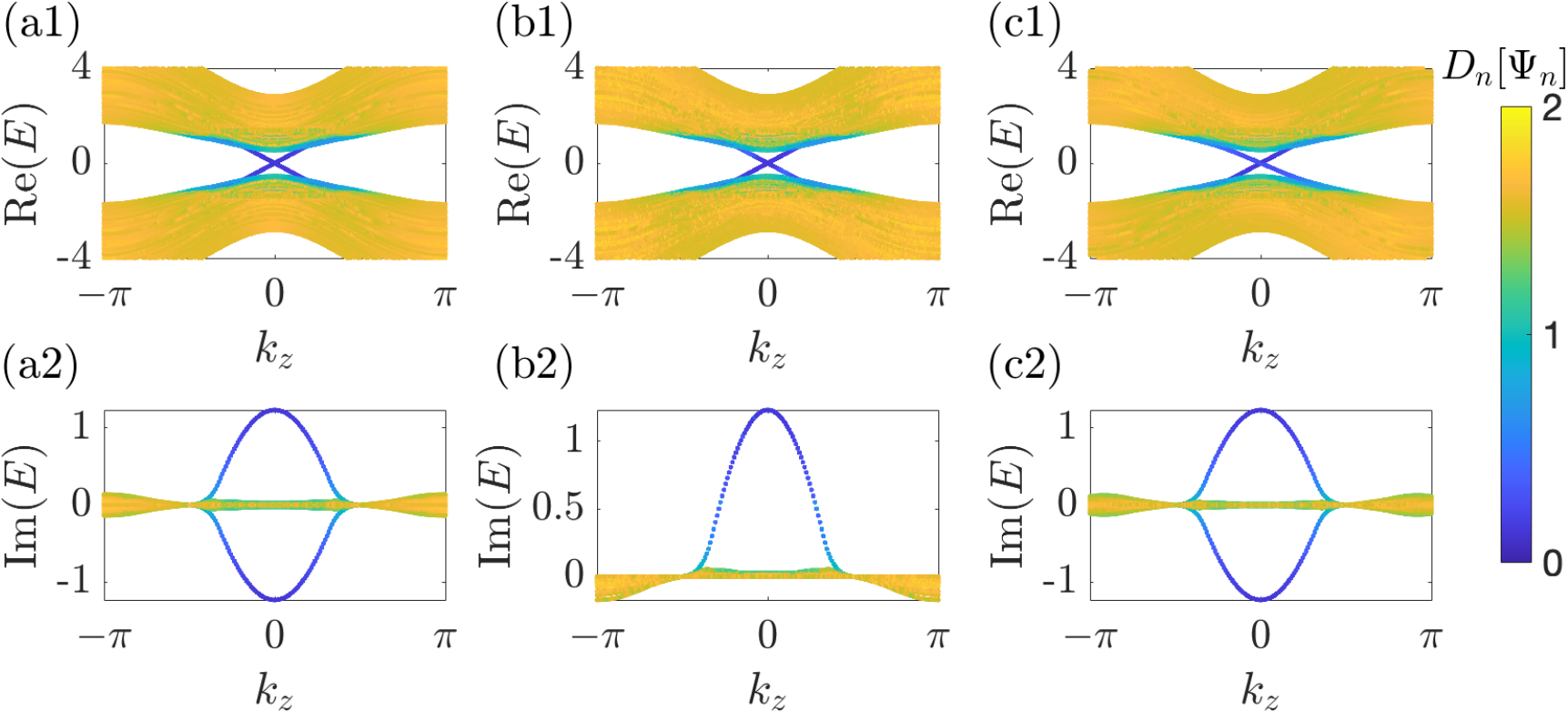}
\caption{
The spectrum of the system with non-Hermitian hinges under $z$-PBCs and $(x,y)$-OBCs. The spectrum is marked by different colors according to the fractal dimension $D_n[\Psi_n]=-\ln[\sum_{x,y}\vert\psi_{n}(x,y)\vert^4]/\ln\sqrt{L_x L_y}$, where $\Psi_n$ is an eigenstate, $\psi_{n}(x,y)$ is the wave amplitude of $\Psi_n$ at position $(x,y)$. The parameters are $m_1=-1.5$, $m_2=-0.5$, $N_x=N_y=8$, (a) $\gamma_A=\gamma_C=-\gamma_B=-\gamma_D=0.75$, (b) $\gamma_A=\gamma_B=\gamma_C=\gamma_D=0.75$, and (c) $\gamma_A=-\gamma_D=0.75$, $\gamma_B=\gamma_D=0$.}
\label{compare2}
\end{figure}

Note that the purpose of us deriving the effective boundary Hamiltonian $H_\text{eff}^\text{(nH-D)}$ is to describe the hinge states which appear only in a certain region of $k_z$ centered at $k_z=0$. Hence we need to choose the range of $k_z$ properly for $H_\text{eff}^\text{(nH-D)}$. The width of the bulk band gap depends on $m_1$ and $m_2$. Meanwhile, given that the non-Hermiticity is only introduced along the hinges and the bulk states hardly occupy the lattice sites along the hinges, the bulk states roughly possess real eigenenergies. Therefore the eigenenergies of the hinge states are complex when those eigenenergies lie within the bulk band gap, and the hinge states disappear when their eigenenergies approximately become real, as can be seen in Fig.~\ref{compare2}(a2/b2/c2). Because $H_\text{eff}^\text{(nH-D)}$ in Eq.~\eqref{eq:hinge} becomes Hermitian and takes real eigenvalues when $k_z=\pm\pi/2$, an effective description of the hinge states shall be provided by $H_\text{eff}^\text{(nH-D)}$ only with $k_z\in[-\pi/2,\pi/2]$. Although $k_z$ is not a good quantum number under OBCs, we may associate each OBC eigenenergy with a specific $k_z$ through the generalized Brillouin zone~\citep{yao2018edge,yokomizo2019non}. Note that each OBC eigenenergy can also be obtained by the spectral evolution during a PBC-OBC transition~\citep{lee2019anatomy}.

\section*{S6. The trade-off between the effective coupling $\Delta$ and the non-normality of topological boundary Hamiltonian}

According to our mechanism, the effective couplings can only be no effect in the following two situations: one is when the non-normality of the boundary Hamiltonian matrix is fixed and the effective couplings are small enough; the other is when the value of the effective couplings (i.e., the size of the bulk) is fixed and the non-normality of the boundary Hamiltonian matrix is small enough. In the main text, we have showed that boundary point-gap topological phases survive under full OBCs when the effective couplings play a significant role. Here, to further support our statement that effective couplings are a mechanism causing boundary point-gap topological phases under OBCs, we will investigate the case of the effective coupling work.

Recall that the irreciprocal hopping of the Hamiltonian can enhance the non-normality of the Hamiltonian matrix and exhibit NHSE under OBCs. For NHSE, the localization of the eigenstates depends on the system's size~\citep{nakai2024topological}: the larger the system's size, the more localized the eigenstates become (i.e., the greater the non-normality of the Hamiltonian matrix becomes).

\begin{figure}[hbt]
\includegraphics[width=1\linewidth]{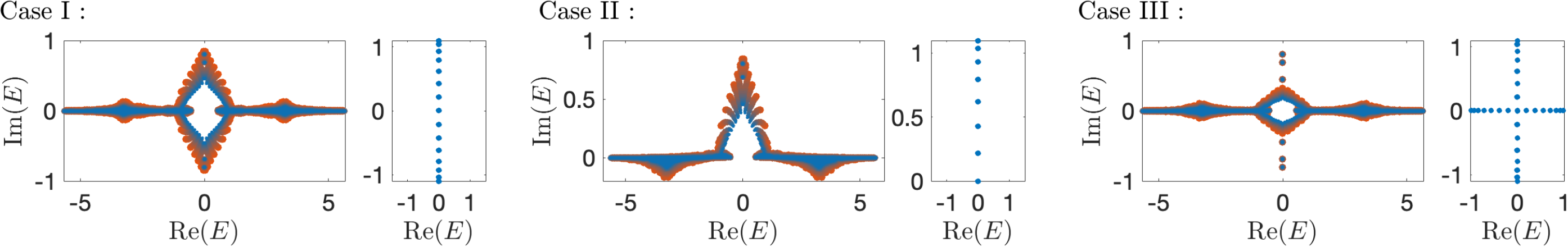}
\caption{The left panel : The spectral flow for the three cases in Table.~I in the main text. The orange-blue curves are the spectral flow from $N_x=N_y=8$ to $N_x=N_y=24$.
The right panel: The spectrum of the effective boundary Hamiltonian where the effective couplings are no effect. The other parameters are $m_1=-1.5$, $m_2=-0.5$, and $\gamma=0.75$.}
\label{supp4}
\end{figure}

We consider decreasing the non-normalities of the boundary Hamiltonians by reducing the boundaries' sizes, and decreasing the effective couplings by enlarging the bulk's size. The relevant results are shown in Fig.~\ref{supp4}. Based on this figure, one can argue that the loop of the point-gap becomes smaller as the value of the effective couplings decreases, which is induced by the increase of the bulk's size. Note that the results in thermodynamic limit cannot be numerically calculated, but one can infer from Fig.~\ref{supp4} that when the bulk is large enough (the thermodynamic limit), the boundaries cannot be sensitive to the effective couplings, preventing the persistence of the point-gap under full OBCs.

{
\section*{S7. The mechanism under generalized boundary conditions}
}

{
In our work, the mechanism of the phase transition fundamentally requires the system to host a topological boundary ($d\geq 1$) whose Hamiltonian is non-normal. However, large boundary parameters may potentially generate non-topological localized bound states~\citep{verma2024topological,zhang2024ptsymmetric}, posing significant challenges to the prerequisites of our mechanism. Furthermore, large boundary parameters could also induce bound states on the topological boundary. In such a case, the Hilbert subspace corresponding to the topological boundary Hamiltonian may host bound states and residual topological boundary states. The behavior of these residual topological states can be predicted by our mechanism and they still exhibit SFL.}

{To study whether our mechanism is still in effect under generalized boundary conditions (GBCs), we consider the non-Hermitian BBH model mentioned in main text, i.e.,
\begin{equation}
\begin{aligned}
H^\text{(nH-BBH)} =& \sum_{j=1}^{L_y}\left(\sum_{i=1,3,\cdots}^{L_x-1}\left(t^\prime a_{i,j}^\dagger a_{i+1,j}+\text{H.c.}\right) + \sum_{i=2,4,\cdots}^{L_x-2}\left(t_x a_{i,j}^\dagger a_{i+1,j}+\text{H.c.}\right)\right) \\
&-\sum_{i=1,3}^{L_x-1}\left(\sum_{j=1,3,\cdots}^{L_y-1}\left(t^\prime a_{i,j}^\dagger a_{i,j+1}+\text{H.c.}\right) + \sum_{j=2,4,\cdots}^{L_y-2}\left(t_y a_{i,j}^\dagger a_{i,j+1}+\text{H.c.}\right)\right) \\
&+\sum_{i=2,4}^{L_x-2}\left(\sum_{j=1,3,\cdots}^{L_y-1}\left(t^\prime a_{i,j}^\dagger a_{i,j+1}+\text{H.c.}\right) + \sum_{j=2,4,\cdots}^{L_y-2}\left(t_y a_{i,j}^\dagger a_{i,j+1}+\text{H.c.}\right)\right)\\
&+\sum_{i=2,4,\cdots}^{L_x-2}\left((t_x+\gamma) a_{i,L_y}^\dagger a_{i+1,L_y}+(t_x-\gamma) a_{i+1,L_y}^\dagger a_{i,L_y}\right)\\
&+\sum_{i=2,4,\cdots}^{L_x-2}\left((t_x-\gamma) a_{i,1}^\dagger a_{i+1,1}+(t_x+\gamma) a_{i+1,1}^\dagger a_{i,1}\right).
\end{aligned}
\end{equation}
The Hamiltonian of this model under GBCs reads
\begin{equation}
H = H^\text{(nH-BBH)} + H_\text{GBC},
\end{equation}
where $H_\text{nH-BBH}$ is the Hamiltonian of the model and
\begin{equation}
H_\text{GBC} = t_x e^{\beta} \sum_{i=1}^{L_y}\left(a^\dagger_{1,i} a_{L_x,i}+a^\dagger_{L_x,i} a_{1,i}\right)-
t_y e^{\alpha} \sum_{j=1,3,\cdots}^{L_x-1} \left(a^\dagger_{j,1} a_{j,L_y}+a^\dagger_{j,L_y} a_{j,1}\right)+
t_y e^{\alpha} \sum_{j=2,4,\cdots}^{L_x} \left(a^\dagger_{j,1} a_{j,L_y}+a^\dagger_{j,L_y} a_{j,1}\right),
\end{equation}
with the y-axis boundary parameter being modulated by $e^{\alpha}$ and the x-axis boundary parameter being modulated by $e^{\beta}$. Consider the model with the same parameters as in the main text under two distinct boundary conditions: (i) OBCs along the x-axis ($\beta\rightarrow\infty$) and GBCs along the y-axis, and (ii) OBCs along the y-axis ($\alpha\rightarrow\infty$) and GBCs along the x-axis.
}

\begin{figure}[htb]
\includegraphics[width=1.0\linewidth]{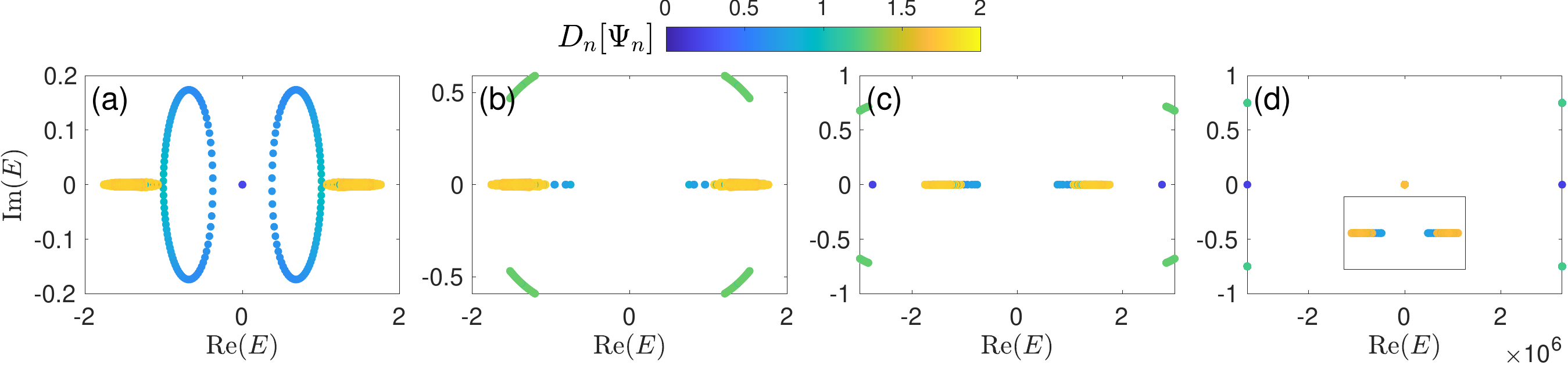}
\caption{{ The 2D non-Hermitian BBH model under OBCs along the x-axis and GBCs along the y-axis. The spectra are marked by different colors according to the fractal dimension. (a) is case under x-OBC and y-GBC with negative boundary parameter $\alpha$, which is close to x-OBC and y-OBC. (b) is case under x-OBC and y-PBC. (c) and (d) are cases under x-OBC and y-GBC with large boundary parameter $\alpha$, where the inset is Re$(E)\in [-2,2]$. We fix the system size $N_x=40$ and $N_y=10$, and choose the same parameter as Fig.~2 in the main text. The other parameters are (a) $\alpha=-30$, (b) $\alpha=0$, (c) $\alpha=1$, and (d) $\alpha=15$.}}
\label{yGBC}
\end{figure}

\begin{figure}[htb]
\includegraphics[width=1.0\linewidth]{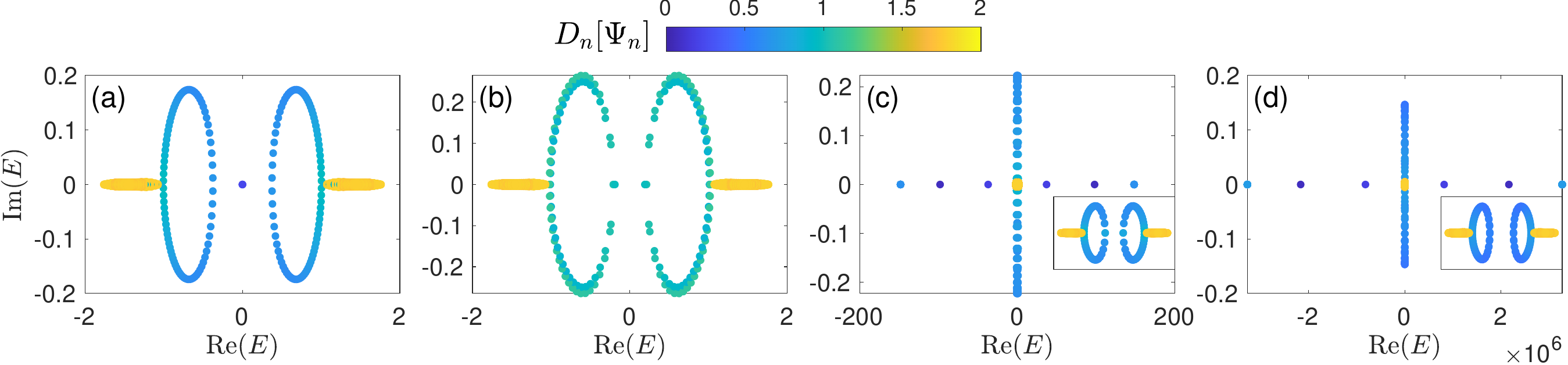}
\caption{{ The 2D non-Hermitian BBH model under OBCs along the y-axis and GBCs along the x-axis. The spectra are marked by different colors according to the fractal dimension. (a) is case under y-OBC and x-GBC with negative boundary parameter $\alpha$, which is close to y-OBC and x-OBC. (b) is case under x-OBC and y-PBC. (c) and (d) are cases under y-OBC and x-GBC with large boundary parameter $\alpha$, where the inset is Re$(E)\in [-2,2]$. We fix the system size $N_x=40$ and $N_y=10$, and choose the same parameter as Fig.~2 in the main text. The other parameters are (a) $\beta=-30$, (b) $\beta=0$, (c) $\beta=5$, and (d) $\beta=15$.}}
\label{xGBC}
\end{figure}

\begin{figure}[htb]
\begin{center}
\includegraphics[width=0.65\linewidth]{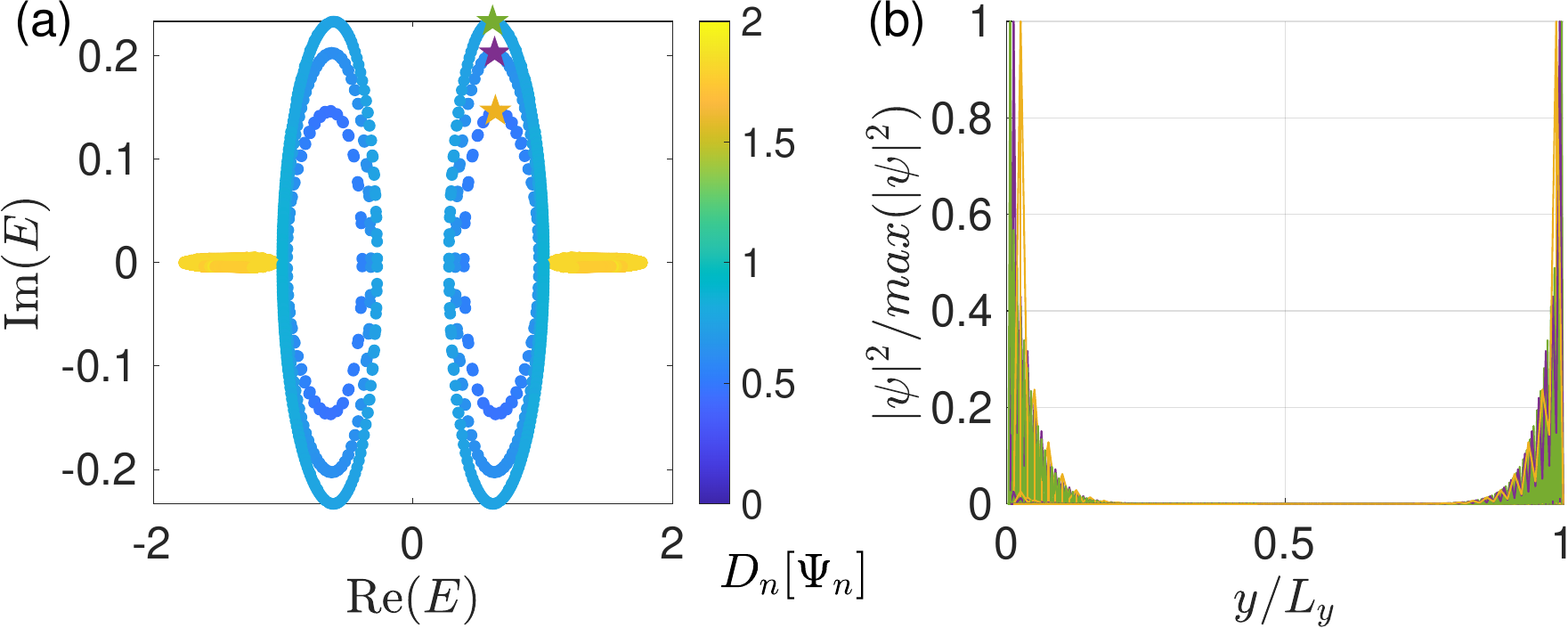}
\caption{{SFL of the first-order topological boundary states in the 2D non-Hermitian BBH model under y-OBC and x-GBC. (a) The spectra Re$(E)\in[-2,2]$ of the model with boundary parameter $\beta=15$ and $N_x=40,80,160$. (b) Rescaled distribution of the boundary states with maximal imaginary energy (marked by the stars in (a)), shows a consistent profile across different $N_x$, along the top and bottom edges ($y = 1$ and $y = L_y$) of the model. The other parameters are same as Fig.~2 in the main text.}}
\label{xGBCsfl}
\end{center}
\end{figure}

{
In Fig.~\ref{yGBC}, we show the spectra of the non-Hermitian BBH model under case (i) with different boundary parameters. Fig.~\ref{yGBC}(b) shows that the model only has the topological boundaries in x-axis under this boundary condition. Because the boundary Hamiltonian corresponding to these topological boundary are Hermitian, the spectrum of these topological states forms a line construct and lies in the real-axis. The spectra escaped from the bulk spectrum originate from two coupled non-Hermitian SSH chains whose couple strength is described by a large boundary parameter, where the corresponding eigenstates distribute only at one of the top and bottom edges. Fig.~\ref{yGBC}(c) and (d) are the large boundary parameter cases. As the value of the boundary parameter increases, the spectra become far and far away from each others. This is out of the scope of our mechanism.}

{
 In Fig.~\ref{xGBC}, we show the spectrum of the non-Hermitian BBH model under case (ii). In this case, the model preserves the y-axis topological boundaries, where the Hamiltonians corresponding to these boundaries are non-Hermitian and their matrices are non-normal. This is in the scope of our mechanism. As shown in Figs.~\ref{xGBC} (c) and (d), the large boundary parameter induces the bound corner states whose eigenenergies are far away from the bulk spectrum, and the scale-free topological states still exist. We also display the SFL of the first-order boundary states in the 2D non-Hermitian BBH model under y-OBC and x-GBC, see Fig.~\ref{xGBCsfl}. }

{
\section*{S8. The effective regime for the mechanism}
}
{
We begin by recalling the Bauer-Fike theorem and the definition of pseudospectrum. Next, we give the relation between the product of non-normality and matrix norm of perturbation, and the pseudospectrum. Finally, we rigorously characterize the parameter regime where our mechanism becomes dominant. \\
\textbf{Bauer-Fike theorem:}\\
Let $A\in M_n$ be diagonalizable, and suppose that $A=V\Lambda V^{-1}$, in which $V$ is nonsingular and $\Lambda$ is diagonal. Let $E\in M_n$ and let $\Vert\cdot\Vert$ be a matrix norm on $M_n$ that is induced by an absolute norm on $\mathbb{C}^n$. if $\mu$ is an eigenvalue of the perturbed matrix $A+E$, then there exists some eigenvalue $\lambda_i$ of $A$ such that:
\begin{equation}
|\mu-\lambda_i|\leq \kappa(V)\Vert E\Vert,
\end{equation}
where $\kappa(\cdot)$ is the condition number.\\
\textbf{Definition of the pseudospectrum:}\\
The $\varepsilon-$pseudospectrum of $A$, denoted $\Lambda_\varepsilon(A)$, is defined as the union of perturbed spectra:
\begin{equation}
\Lambda_\varepsilon(A) = \bigcup_{\Vert E\Vert\leq \varepsilon} \sigma(A+E),
\end{equation}
where $\sigma(A+E)$ is the set of eigenvalues of $A+E$.
}
{
Now, we explain the relation between the product of non-normality and matrix norm of
perturbation, and the pseudospectrum. By the Bauer-Fike inequality, every $z\in \Lambda_\varepsilon(A)$ must lie within a distance $\kappa(V)\varepsilon$ from at least one original eigenvalue. Consequently, the pseudospectrum is contained within the union of disks centered at $\lambda_i$ with radii $\kappa(V)\varepsilon$:
\begin{equation}
\Lambda_\varepsilon(A) \subset \bigcup_{i=1}^n D\left(\lambda_i,\kappa(V)\varepsilon\right).
\end{equation}
This suggests that the product of non-normality and matrix norm of perturbation directly impacts the pseudospectrum.}

{
Then, we characterize the parameter regime where our mechanism becomes dominant. For convenience, we fix the non-normality $\kappa(V)$ but vary the perturbation norm $\Vert E\Vert$ in the following discussion.
The behavior of the spectrum under perturbations can be systematically analyzed through the product $\kappa(V)\Vert E\Vert$, which governs the maximum eigenvalue deviation via the Bauer-Fike Theorem:
\begin{equation}
\text{the maximum eigenvalue deviation} \leq \kappa(V)\Vert E\Vert.
\end{equation}
Analyzing the disks of radius $\kappa(V)\varepsilon$ and the minimal eigenvalue spacing $\delta=\text{min}_{i\neq j}|\lambda_i-\lambda_j|$ leads to three distinct regimes:
\begin{itemize}
\item[a.] Stability regime ($\kappa(V)\Vert E\Vert\ll \delta$ )
\\
Eigenvalues shift linearly with $\Vert E\Vert$, remaining near their original positions. The pseudospectrum $\Lambda_\varepsilon(A)$ consists of isolated disks of radius $\kappa(V)\varepsilon$, reflecting stability.
\item[b.] Transition regime ($\kappa(V)\Vert E\Vert\sim \delta$ )
\\
Pseudospectral disks $D(\lambda_i,\kappa(V)\varepsilon)$ begin overlapping, leading to geometric merging of eigenvalue neighborhoods. This triggers the eigenvalue migration and topological changes in the spectrum.
\item[c.] Instability regime ($\kappa(V)\Vert E\Vert\gg \delta$ )
\\
The pseudospectrum covers large regions of the complex plane, and eigenvalues may deviate catastrophically from their original positions.
\end{itemize}
Here, we define that the critical perturbation threshold for pseudospectral merging is $\Vert E\Vert_\text{critical} = \delta/\kappa(V)$.
}
{
Based on the above discussion, the effective regime for this mechanism satisfies the inequality $\kappa(V)\Vert E\Vert\geq\delta$.
}

\begin{figure}[h]
\begin{center}
\includegraphics[width=0.9\linewidth]{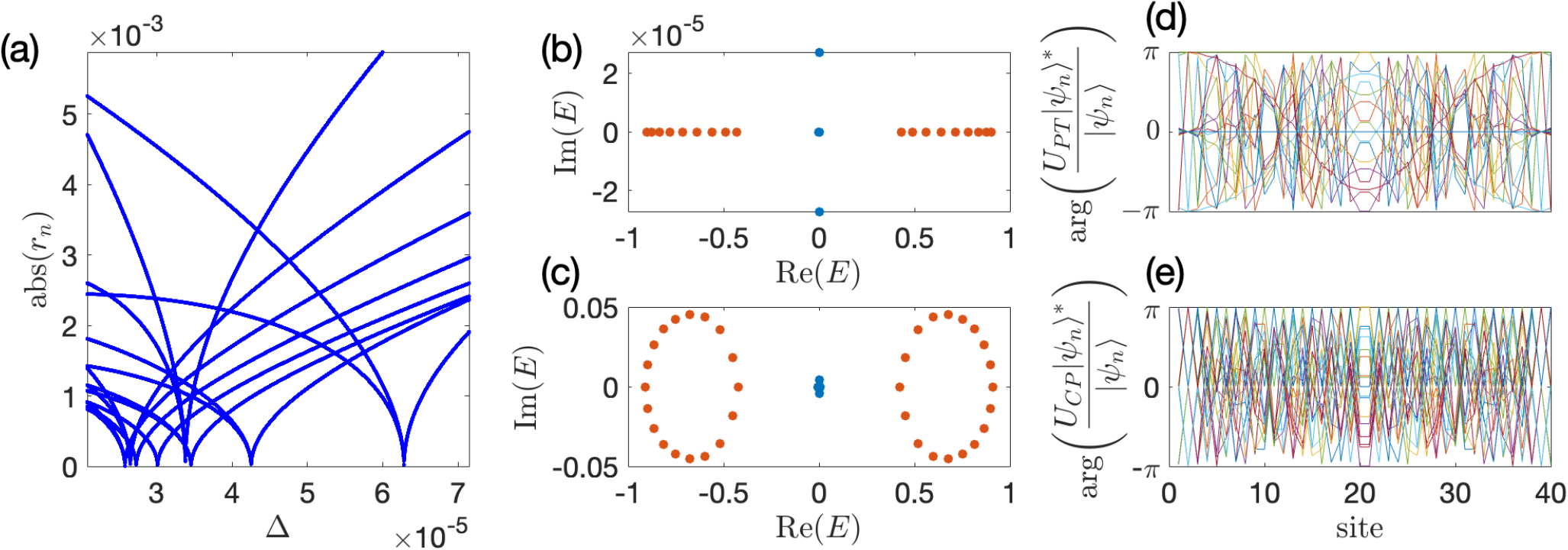}
\caption{{The phase rigidity and the symmetry analysis for the effective boundary Hamiltonian. (a) The magnitude of the phase rigidity $r_n$ for different eigenstates as a function of effective coupling $\Delta$. (b) and (c) are the eigenspectrum for the Hamiltonian with $\Delta = 10^{-8}$ and $\Delta = 10^{-4}$, respectively. (d) is the phase difference between $\ket{\psi_n}$ and $U_\text{PT}\ket{\psi_n}^\ast$ for the effective coupling $\Delta = 10^{-4}$. (e) is the phase difference between $\ket{\psi_n}$ and $U_\text{CP}\ket{\psi_n}^\ast$ for the effective coupling $\Delta = 10^{-4}$.
The other parameters are $t=1$, $\gamma_+=-\gamma_-=0.75$, $t^\prime=0.25$, and $N_x = 10$.}}
\label{fig:symAna}
\end{center}
\end{figure}

{
\section*{S9. The relation between exceptional points and the scale-free localization behavior}
}
{The scale-free localization (SFL) behavior is related to exceptional points (EPs), but not necessarily to PT-symmetry or other symmetries (pseudo-Hermiticity, CP and pseudo-chirality). Indeed, during the phase transition between line-spectrum to loop-spectrum, the eigenspectrum splits into a pair of complex conjugate eigenvalues at the beginning, implying that the system Hamiltonian has the topological structure of EPs. Note that the line-spectrum is not necessarily purely real, and the generation of EPs is only a necessary condition for the symmetry transition. The effective boundary Hamiltonian of the model, presented in Eq. 11 in the main text, preserves both a PT-symmetry and a CP-symmetry. The SFL behavior in this model happens when the pure real spectrum splits into a pair of complex conjugate eigenvalues, which implies that the SFL behavior in this model is related to the transition between PT-symmetry exact and broken phases. To further confirm this, we consider the effective boundary Hamiltonian in Eq.~(11) in the main text. Then we use phase rigidity,
\begin{equation}
r_n = \frac{\bra{\psi_n^L}\psi_n^R\rangle}{\langle\psi_n^R\ket{\psi_n^R}},
\end{equation}
as a measure of the extent to which the wave functions mix around an EP through the deviation of
the normalization from unity, where $\ket{\psi_n^{L(R)}}$ is the eigenstates of the model with the superscript $L$ $(R)$ denoting the left (right) eigenstate. Meanwhile, we analyze the eigenstate symmetry of the model
from the line-spectrum phase to the loop-spectrum phase, where the EPs emerge during this
transition. From Fig.~\ref{fig:symAna}(a), one can see that the magnitude of the phase rigidity has $8$ zero points, suggesting that EPs emerge. Figs.~\ref{fig:symAna}(b) and (c) show that the pure real spectrum splits into a pair of complex conjugate eigenvalues. The line spectrum with real eigenenergies indicates the PT-symmetry exact phase. On the other hand, Figs.~\ref{fig:symAna}(d) and (e) show that the eigenstates in the loop spectrum do not preserve the PT-symmetry and CP-symmetry because $\ket{\psi_n}$ and $U_\text{PT(CP)}\ket{\psi_n}^\ast$ have a position-dependent phase difference, implying that they are different states and SFL behavior is related to the symmetry broken phase in this model.}

\bibliography{refs_BMBI}